\documentclass[useAMS,usenatbib]{mn2e}

\newcommand{\gtsimeq}{\raisebox{-0.6ex}{$\,\stackrel
        {\raisebox{-.2ex}{$\textstyle >$}}{\sim}\,$}}

\begin{document}

\title[Implications for unified schemes from the quasar fraction and 
emission-line luminosities]{Implications for unified schemes from the 
quasar fraction and emission-line luminosities in radio-selected samples}
\author[J.A.Grimes, S.Rawlings \& C.J.Willott]
  {Jennifer A.~Grimes,$^{1\star}$ 
  Steve Rawlings$^1$ and 
  Chris J.~Willott$^{1,2}$ \\ 
  $^1$University ~of ~Oxford, Astrophysics, Keble ~Road, Oxford, 
OX1 ~3RH, UK \\
  $^2$Herzberg Institute of Astrophysics, National Research Council, 5071 West 
Saanich Road, Victoria, BC, V9E 2E7, Canada\\}

\maketitle

\begin{abstract}
\noindent
We use a principal components analysis of radio-selected (3CRR, 6CE and 7CRS)
AGN datasets to
define two parameters related to low-frequency ($151$ MHz) 
radio luminosity $L_{151}$ and [OIII]
luminosity $L_{{\rm [OIII]}}$: a parameter $\alpha$ encoding the 
$L_{151} - L_{{\rm [OIII]}}$
correlation and a parameter $\beta$ encoding scatter about this
correlation. We describe methods for constructing generalized luminosity
functions (GLFs) based on $\alpha, \beta$, redshift and schemes for unifying
quasars and radio galaxies. These GLFs can be used to generate radio 
luminosity functions (RLFs) which improve on those of 
\citet{wrlf}, mostly because they incorporate scatter and are therefore 
much smoother.

Luminosity-dependent unified schemes (e.g. a receding-torus scheme) have
been invoked to explain the low quasar to radio galaxy fraction
at low $\alpha$ and the differences in emission-line luminosities of 
radio quasars and radio galaxies. 
With the constraints of the 3CRR, 6CE and 7CRS datasets and radio 
source counts, our GLF approach was used to determine whether a  
receding-torus-like scheme is required if
there are two populations of radio sources:
one at low $\alpha$,
consisting of `starved AGN'; the other at high $\alpha$
consisting of `Eddington-tuned AGN'.
Because of the overlap between these two populations and the effects of the 
$\beta$ parameter, schemes with or
without a receding torus can produce a low quasar fraction 
at low $\alpha$ and differences in [OIII] luminosity between 
radio galaxies and quasars.
The receding torus may be a physical process important in one or more  
populations of radio sources, but this is not yet proved either by the
quasar fraction or the emission-line properties of radio-selected samples. 
\end{abstract}

\begin{keywords}
 galaxies: active -- quasars: general -- galaxies: evolution
\end{keywords}

\footnotetext{$^{\star}$Email: jag@astro.ox.ac.uk}

\section{Introduction}
\label{sec:intro}

It has been recognised for some time that there is a strong positive 
correlation between the extended-radio luminosities and narrow-emission-line 
luminosities of 3C radio sources \citep{bh, retal, mcarthy}, and by 
combining the 3C sample with the 7C Redshift Survey (7CRS), \citet{wem}
found that these correlations were not primarily due to redshift.
This suggests that the sources of narrow-line emission and radio jets are 
linked,
with explanations for this link ranging from the effects
of environment \citep{dp93}, to jet-disk symbiosis \citep{rs91, fb}.

Unified schemes for radio galaxies and radio quasars propose that they are
the same objects viewed at different angles between
their radio axis and their line of sight.
An obscuring torus is invoked to hide the nucleus at large angles to the jet 
axis i.e. in radio galaxies.
Emission from narrow-line regions is believed to be broadly independent of 
the jet-axis orientation, as it is emitted from a region larger than 
the extent of the torus. 
The opening angle of the torus 
$\Theta_{\rm trans}$ marks the transition
from the object being viewed as a quasar to a radio galaxy.
We therefore expect for simple, arguably naive, unified schemes, 
where $\Theta_{\rm trans}$
is constant, that the distribution of emission-line luminosities should 
be similar
for radio quasars and radio galaxies. \citet{jb} found that [OIII] luminosities
of radio quasars are $\sim 5-10$ times more luminous than radio galaxies. 
In similar comparisons of radio quasars and radio galaxies,
\citet{hbf} found no difference using [OII], and 
\citet{jr} found no difference in [OIII] at high redshift.
In summary, some studies comparing the narrow
emission line strengths of radio quasars and radio galaxies 
are seemingly in agreement with the 
predictions of the simplest unified schemes, whilst others are not.

These seemingly contradictory,
results can perhaps be understood in the context of a `receding torus' model 
for AGN.
This model \citep{law, hgd} proposes
that the inner radius $r$ of the obscuring torus is
determined by the radius at which dust sublimates, scaling as 
$r \propto L_{\rm phot}^{0.5}$.
Assuming that the half-height $h$ of the torus is independent of 
$L_{\rm phot}$, then the half-opening-angle of the 
torus, $\theta = \tan^{-1} r / h$, will
be larger in the higher-$L_{\rm phot}$ objects.
This means that more luminous objects, with higher
$L_{\rm phot}$, are more likely to be viewed as quasars so that 
quasar fraction rises systematically with $L_{\rm phot}$. 
It also means that
orientation-independent quantities which scale with $L_{\rm phot}$ will
be higher on average for quasars than for radio galaxies.
\citet{law} found that the fraction of narrow-line objects in the
low-frequency-selected 3CR sample decreased with increasing radio luminosity,
and that narrow-line objects have weaker [OIII] at a fixed radio luminosity.
He argued that
this was inconsistent with the simplest unified schemes but that it
could be explained by a cone-angle dependence on luminosity, e.g 
a receding-torus model.
\citet{simp} argued that 
the [OIII] emission line is a much better 
indicator of $L_{\rm phot}$ than [OII], explaining why differences between
quasars and radio galaxies are more obvious in [OIII] than [OII].
\citet{sconf} revisited the arguments of \citet{simp} and corrected a small
error (compare Fig.~2 of \citealp{sconf} with Fig.~4 of \citealp{simp}).
He concluded that, allowing for a spread in $h$, the receding torus model
predicts that radio quasars should be a factor of a few brighter in
$L_{\rm phot}$ than radio galaxies in samples exhibiting a wide range of 
quasar fractions.

\citet{wsubmm} found that radio quasars have higher submillimetre
luminosities by a factor of $\sim 4$ than radio galaxies of the same radio
luminosity and redshift, a factor which cannot be reduced below $\sim 2$
by accounting for possible synchrotron contamination. 
This result supports
the idea that the simplest unified schemes, where $\Theta_{{\rm trans}}$ is 
constant, are not an adequate description
of the relationship between radio galaxies and quasars: submillimetre 
emission comes from cool dust grains in optically thin regions, and therefore
radiates isotropically; moreover, such emission could not be obscured 
by a torus even if it were emitted from regions close to the nucleus.
Submillimetre luminosity is therefore an orientation-independent quantity
which might scale closely with $L_{\rm phot}$ either because the cool dust is
heated directly by the quasar or because it is heated by a starburst whose
luminosity scales with $L_{\rm phot}$. We conclude that the submillimetre
study of \citet{wsubmm} is in quantitative agreement with
the receding torus model of \citet{sconf}.

Other arguments for a receding-torus-like model include the evidence 
that the fraction of lightly reddened $z \sim 1$ 3C quasars decreases with 
increasing radio luminosity in agreement with the higher fraction of
lines of sight expected to graze the torus at lower $L_{\rm phot}$
(at fixed $h$) in the receding torus 
model (\citealp*{hgd}; \citealp{srl}). 
Also the near-infrared spectral indices of quasars from the 3CR 
sample are correlated with luminosity, whereas the optical spectral 
indices are uncorrelated with the quasar luminosity or orientation, so that 
the strength of the `big red bump', relative to the ionizing continuum, 
appears to be less in the more luminous objects \citep{sr,sconf}.

However, there is significant scatter in the relationship between radio and
emission-line luminosities and a receding torus is not
the only way of explaining the differences in the emission-line properties
of radio quasars and radio galaxies or the luminosity-dependence of the
quasar fraction (the fraction of objects that show broad emission lines).
Two-population radio luminosity functions (RLFs) have been used to 
provide a best-fit to the 3CRR, 6CE and 7CRS radio source redshift surveys 
and radio source counts \citep{wrlf}.
It is possible that a two-population model with a simple unified 
scheme in one population, combined with the effects of scatter, could mimic 
the effects of a receding torus in producing both emission-line differences 
between radio galaxies and radio quasars and the gradual rise in quasar 
fraction with emission-line luminosity.
\citet{wqf} found a drop in the quasar fraction of AGN at low
luminosity, postulating that their results are consistent with either the 
emergence of a second population of low-luminosity radio sources, which 
lack a well-fed nucleus, or a gradual decrease
in $\Theta_{{\rm trans}}$ with decreasing radio luminosity.

Traditionally, radio galaxies have been divided into two populations, 
FRI and FRII, \citep{fr}, based on 
radio structure but correlated with radio luminosity such that above 
$\log_{10}(L_{151} / {\rm W\,Hz^{-1}\,sr^{-1}}) \sim 25$,
objects are FRII, whereas below that critical
$L_{151}$, they are FRI. \citet{ccc02} found that the FRII population is 
inhomogeneous and that not all of them can be unified with quasars. 
They find a low-radiative-efficiency accretion, weak or absent broad-line 
emission and a lack of a significant nuclear absorbing structure for weak-jet, 
low-ionisation narrow-line galaxies. For broad-line objects and 
obscured high-ionization narrow-line galaxies they see or infer 
intense ionizing emission, powerful jets and a torus-like absorber.
In low-radio-frequency-selected samples, the fraction of objects with 
observed broad lines changes rapidly from $\sim 0.4$ 
above $\log_{10}(L_{151} /{\rm W\,Hz^{-1}\,sr^{-1}}) \sim 26.5$ to 
$\sim 0.1$ below this.
The less luminous population from \citet{wrlf} is composed 
of FRIs and FRIIs with weak/absent emission lines and their more 
luminous population of strong-emission-line FRII radio galaxies and quasars. 
A two-population model like this is motivated by the fact that the 
presence or absence of a quasar nucleus (as shown by emission-line strength) 
seems likely to be connected to the properties of a central engine of the 
radio source. \citet{ka97}, and others before them, have argued that 
radio structure is influenced by the environment. It seems likely that a 
high jet power is necessary for a highly collimated non-dissipative jet, thus 
all high luminosity sources are FRII, but as lower jet powers are reached, a 
jet is more likely to disrupt resulting in a FRI source.
The exact environmental density and homogeneity would determine the 
radio luminosity threshold of FRI/FRII
within the lower luminosity population. \citet{lo96} found that the
FRI/FRII division is proportional to the square of the optical 
luminosity of the host galaxy, which is plausibly related to such 
environmental effects, although
other causes have been proposed.
This simple picture is shown to be incomplete by the discovery by
\citet{br} of an optically powerful quasar with FRI radio structure.
It also seems that there is evidence that some FRI sources in 3C have hidden 
quasar nuclei \citep{cao}, and one clear example of a 3CR FRI with broad lines 
(3C386, \citealp{s96}) is known.
In this paper we follow \citet{hl}, \citet{laing} and \citet{jw}
in assuming that the FRI/FRII division is strongly influenced by
environmental effects and is much less fundamental than a division 
based on accretion properties of the central engines.

It is well known (e.g. \citealt{dp90}) that the evolution in comoving space density
of radio galaxies depends on both $L_{151}$ and $z$. In two-population
models this can be expressed as differential density evolution of
these populations \citep{wrlf}. \citet{sb} have criticized this
approach on the basis of the artefacts it produces when the populations
join.

The primary motivation of this work was to investigate in a quantitative manner
the differences between the emission-line properties of radio-loud
quasars and radio galaxies alongside the luminosity dependence of the 
quasar fraction.
In this paper we will quantify the three-dimensional distribution in
[OIII] emission-line luminosity $L_{{\rm [OIII]}}$, $151$ MHz luminosity $L_{151}$ and redshift
$z$ using a new method applied to complete samples.
A principal components analysis (PCA) is used to find the axis which causes the most
differentiation between the objects, and it is possible that it can identify parameters which
are more physically meaningful than the attributes of each data point (radio and
emission-line luminosity) used in the PCA.
We carry out a PCA and
define a generalized luminosity function (GLF) in this new parameter space.
The GLF is then used to make proper comparisons between receding-torus and two-population models,
allowing for cosmic evolution of the populations.

The paper is organized as follows. In 
Sec.~\ref{sec:data} the data are described.
In Sec.~\ref{sec:pca}, the principal
components analysis is described. 
In Sec.~\ref{sec:model}, simple one-population GLFs are described to 
enable a comparison between unified schemes with and without a receding torus.
The effects of using  two-population 
GLFs are presented in Sec.~\ref{sec:2pop}, and in Sec.~\ref{sec:discussion}
we compare our results with previous studies and
discuss possible meanings for the new parameters found by the PCA.
We assume throughout that
${\rm H_0} = 70\,{\rm km\,s^{-1}\,Mpc^{-1}}, \Omega_{\rm M} = 0.3$ and 
$\Omega_{\Lambda} = 0.7$.

\section{Data}
\label{sec:data}

\subsection{Classification of radio sources}
\label{sec:class}
Most theories
postulate the existence of a central accreting
supermassive black hole in all radio sources, so any division between radio quasars and
radio galaxies is bound to be problematic. Operationally, however, the
division seems reasonably clean. Previous prescriptions for
making this division \citep[e.g.][]{jr, wqrlf}
have defined radio quasars as objects whose integrated optical
luminosities are
dominated by point sources (rather than the host galaxy). For most other radio sources, defined as
radio galaxies, there is no direct detection of a compact nuclear source or
broad emission lines, although of course both could be present but
obscured. Following \cite{hl} and
\cite{laing}, most authors tend to
sub-divide radio galaxies into those with high-excitation
narrow-line spectra (HEGs) and those with low-excitation narrow-line
spectra (LEGs). We follow \cite{jr} in defining LEGs as objects
with (rest-frame) [OIII] equivalent widths of $< 10 ~ \rm \AA$, [OII]/[OIII]
ratios $>1$, or both.

Amongst HEGs, there are a number of cases where
classification as a radio quasar is arguably more natural than
classification as a radio galaxy. This
number has tended to increase as observational methods have improved,
and become more varied and sophisticated. Such objects, often 
called `weak quasars' can be sub-divided into a number of
distinct categories: (i) objects with unobscured, broad-line
optical nuclei which are insufficiently luminous to outshine the
host galaxy (3C382, \citealp{c99}; 3C390.3, \citealp{perez};
5C7.17, 5C7.118, \citealp{wqrlf}); (ii) objects with lightly veiled
(dust obscuration $A_{\rm V} \sim 1-5$) broad-line nuclei
seen via broad wings on the H$\alpha$ line
(3C33.1, 3C67, 3C268.3, \citealp{laing}; 3C22, 3C41, 
\citealp*{srl}; 3C109, \citealp{c99}; 3C303, \citealp{eh}) 
or via broad Paschen lines (3C184.1, 3C219, 3C223, \citealp*{hgd}),
via broad wings on rest-frame UV lines (3C325, Grimes et al.
in prep), or inferred from nuclear point sources seen in the thermal IR (3C65,
\citealp*{srl}; 3C79, 3C234, \citealp{sww}) or, at high spatial
resolution, with the HST (3C455, \citealp{lehnert});  and (iii)
objects with broad lines apparent only in optically-polarized light,
presumably scattered from a nucleus with $A_{\rm V} \sim 10$
(3C265, \citealp*{dey,srl}; 3C226, \citealp{diSCF}). 
The thermal-IR observations of \citet*{srl} show, nevertheless, 
that material of high ($A_{\rm V} \gg 10$)
visual extinction is needed to hide any bright nuclei in most HEGs.
It is highly plausible (c.f Cyg A, \citealp{ogle}) that some of these 
highly obscured nuclei will eventually show scattered (i.e. polarized)
broad lines in deep spectropolarimetry.

\begin{figure}
\begin{center}
\setlength{\unitlength}{1mm}
\begin{picture}(150,60)
\put(0,-10){\includegraphics{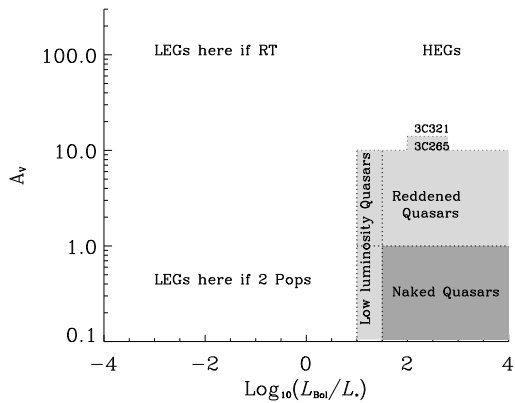}}
\end{picture}
\end{center}
{\caption[junk]{\label{fig:qorg}
{
A cartoon showing an idealized way of classifying the optical
counterparts of radio sources. The visual extinction
($A_{\rm V}$) towards the compact optical nucleus is plotted against the
(unobscured) bolometric luminosity ($L_{\rm Bol}$) of this
nucleus (in units of the luminosity $L_{*}$ at the 
break of the galaxy luminosity function). 
The areas of this plane are 
labelled with the types of object
they are likely to contain. The shaded region shows the location of 
objects defined as radio quasars (which have $M_B < -23$),
with lighter shading illustrating the
location of `weak quasars' (WQ) -- note that this includes
both weak and/or veiled objects and, rather arbitrarily
(see text), includes some quasar nuclei (like 3C265) in which broad lines
are detected only after scattering and excludes some objects (like 3C321, \citealp{c99})
although scattered broad lines are seen. The lower luminosity limit for weak quasars
was derived from the lowest luminosity 7CRS weak quasar.
The location of the LEGs in this
plane is strongly dependent on whether a two-population-like
scheme or a receding-torus-like scheme provides a better physical
description of the radio source population.
}}}
\end{figure}

For the LEGs there seems little doubt that any quasar nuclei
are of comparatively low bolometric luminosity, but whether
or not they are naked (as argued
by \citealp{ccc99}) or buried (as predicted by the receding torus
model) is still open to debate. There is one clear example of
a broad-line, optically-compact nucleus in a LEG (3C386, \citealp{s96,ccc99}). 

With this bewildering variety of ways in which the terms radio quasars and
radio galaxies can become confused, it is clear that any
division into two classes needs to be carefully explained.
Fig.~\ref{fig:qorg}  is a cartoon of a classification scheme based on both the
intrinsic (unobscured) bolometric
luminosity of nucleus and the visual extinction to the nucleus.
It is worth emphasising that how one goes about meaningfully
separating radio quasars and radio galaxies
depends on the nature of the question being asked. Here, as
discussed in detail in Sec. 1,  we are principally interested in comparing 
a two-population-like model in which LEGs have naked, but
very low luminosity, nuclei with  a receding-torus-like model in which
$L_{\rm Bol}$ controls $A_{V}$ so that LEGs necessarily
have high $A_{V}$. The natural division for this problem,
illustrated in Fig.~\ref{fig:qorg}, separates objects with any direct evidence for a
quasar nucleus, veiled or not and weak or not, from those in
which there is no such evidence. In most figures we will
plot weak quasars (WQs), which are significantly
($A_{V} \gtsimeq 1$) veiled and/or low luminosity,
with a separate symbol.
Whether or not one defines objects with broad lines only
in scattered light is a matter of debate because broad lines could, in
theory, be scattered from highly obscured nuclei. Considering 
such objects, we 
somewhat arbitrarily classify 3C226 and 3C265 as weak quasars -- in the latter case 
because a thermal-IR nucleus with $A_{\rm V} \sim 10$ is seen (\citealp{srl})  -- 
and 3C321 as a radio galaxy because 
although scattered broad H$\alpha$ is seen
(\citealp{c99}), transmitted
broad Paschen $\alpha$ is not (\citealp*{hgd}).

\subsection{Complete samples}
\label{sec:samples}
A complete sample consists of every radio source in a certain area of sky
brighter than a specified flux-density limit at the specified selection frequency.
Ideally, all the sources are identified optically with a radio galaxy or quasar, the
redshifts and emission-line luminosities are determined for each source
and the sample sky area and flux-density limits are known.
Radio ($151$ MHz) and emission line ([OIII]) data were used from the
3CRR sample of \citet{lrl}, the 6CE sample (
\citealt{rel}, a revision of the 6C sample of \citealt{eales}) and the 
7CRS sample \citep{wspec}.

The 3CRR sample is selected with $S_{178} \geq 10.9$ Jy
($S_{151} \geq 12.4 $ Jy assuming a spectral index of $0.8$), and with a sky
area of $4.23$ steradians.
Three sources are excluded, (3C231 because the radio emission is due to
a starburst and not an AGN, and the flat-spectrum quasars, 3C345 and 3C454.3
as they are only in the sample because of Doppler boosting), leaving $170$ sources \citep{wrlf}.
Of these, $39$ are classified as quasars, $20$ as weak quasars, and $111$ are radio galaxies,
(see Table \ref{tab:samples} for a summary of all samples).
The radio galaxies consist of $88$ FRIIs and $23$ FRIs, and unless otherwise specified, the term radio
galaxies will henceforth refer to both FRIIs and FRIs.

\begin{table*}
\footnotesize
\begin{center}
\begin{tabular}{llccccccc}
\hline\hline
\multicolumn{1}{c}{Sample}&\multicolumn{1}{c}{Area}&\multicolumn{1}{c}{$S_{151}$ lower limit}&\multicolumn{1}{c}{$S_{151}$ upper limit}&\multicolumn{1}{c}{Q}&\multicolumn{1}{c}{WQ}&\multicolumn{1}{c}{FRII RG}&\multicolumn{1}{c}{FRI RG}&\multicolumn{1}{c}{Total}\\
&\multicolumn{1}{c}{sr}&\multicolumn{1}{c}{Jy}&\multicolumn{1}{c}{Jy}&&&&&\\
\hline
3CRR&$4.23$&$12.4$&-&$39$&$20$&$88$&$23$&$170$\\
6CE&$0.103$&$2.0$&$3.93$&$9$&$0$&$49$&$0$&$58$\\
7CRS(I and II)&$0.013$&$0.5$&-&$23$&$2$&$44$&$5$&$74$\\
\hline
Total&&&&$71$&$22$&$181$&$28$&$302$\\
\hline\hline
\end{tabular}\\
{\caption[Sample table]{\label{tab:samples}{Table of the complete samples used
to constrain the GLFs. 
}}}
\end{center}
\normalsize
\end{table*}

The flux-density limits of the 6CE sample are $2.0 \leq S_{151} \leq 3.93$ Jy, and the sky area is $0.103$ sr,
and $58$ sources are used (as one object 6C1036+3616 is occluded by a bright star and
is excluded without bias), of which $49$ are radio galaxies and $9$ are radio
quasars.

The 7C-I and 7C-II subsamples of the 7CRS are
used, consisting of $37$ and $39$ sources, with $0.0061$ sr and $0.0069$ sr
and $S_{151} \geq 0.51$ Jy and $S_{151} \geq 0.49$ Jy respectively.
Two sources are excluded: $3{\rm C}200$ because it is already included
in the 3C sample and $5{\rm C}7.230$ as it is a flat-spectrum object
which is only in the sample because of Doppler boosting.
Of these objects, $49$ are radio galaxies, $23$ radio quasars and
$2$ weak quasars (5C7.17 and 5C7.118). 

\begin{figure}
\begin{center}
\setlength{\unitlength}{1mm}
\begin{picture}(150,80)
\put(-5,0){\includegraphics{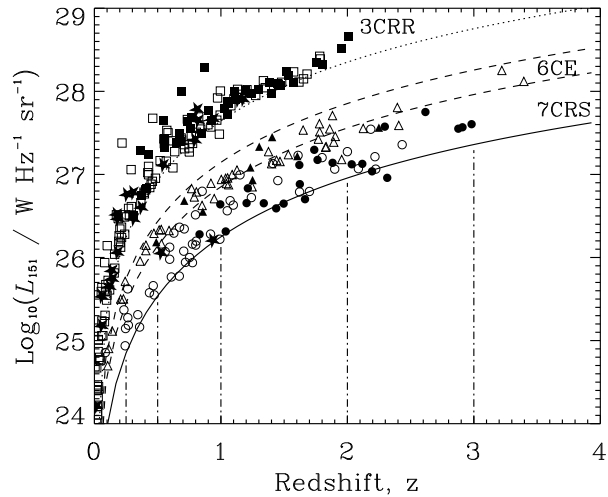}}
\end{picture}
\end{center}
\vspace{-1cm}
{\caption[junk]{\label{fig:pz} 
{The $151$ MHz luminosity $L_{151}$ versus redshift $z$ plane for the 3CRR, 6CE and 7CRS samples
described in Sec.~\ref{sec:samples} and Table \ref{tab:samples}.
The various symbols identify radio quasars and
radio galaxies from different samples: 3CRR quasars (filled squares);
3CRR radio galaxies (open squares); 6CE quasars (filled triangles); 6CE radio galaxies (open triangles); 7CRS quasars (filled circles) and 7CRS radio galaxies (open circles).
Weak Quasars (see Sec.~\ref{sec:class}) are shown by filled stars in both the 3CRR and 7CRS samples.
The dotted line shows the flux-density lower limit for the 3CRR sample, the dashed lines show the upper and lower limits for 6CE and the solid line shows the
lower limit for 7CRS.
The dot-dashed lines mark the $L_{151}$ limits of the 7CRS sample at 
$z = 0.25, 0.5, 1.0, 2.0, 3.0$, as used in Fig.~\ref{fig:lf}.
}}}
\end{figure}

Even coverage of the radio luminosity-redshift plane would provide the
best constraint with which to derive luminosity functions.
However from Fig.~\ref{fig:pz}
we can see that the complete samples described above have a very uneven
distribution on the $L_{151}-z$ plane, because the samples are flux-density limited and quite
small. However, the 3CRR, 6CE and 7CRS samples do have very high spectroscopic
completeness and there are reliable redshift estimates for the small fraction
of objects without spectroscopic redshifts \citep{wrb}.

\subsection{Emission-line data}
The emission-line data for the 6CE sample are given in
\citet{rel} and for the 7CRS sample in \citet{wspec}.
For the 3CRR survey, the emission-line data are taken primarily from \cite{jr},
\citet{hjr}, R.~Laing and J.~Wall
(priv. comm.) following \citet{wem}. 

Since the sources in the samples described above have redshifts in the
range  $0 < z < 4$, it is not possible to measure the flux of the same
emission line in each object with optical spectroscopy and the availability
of near-infrared spectroscopy is limited.
The [OIII] $\lambda 5007$ line is chosen as the second luminosity
in the model because it is an excellent indicator
of the strength of the underlying non-stellar continuum \citep{simp},
and it is probably produced at radii beyond those obscured by the torus.
Note that \citet{wem} used [OII] because it was the line with the 
most measurements.

\begin{figure}
\begin{center}
\setlength{\unitlength}{1mm}
\begin{picture}(150,50)
\put(0,-10){\includegraphics{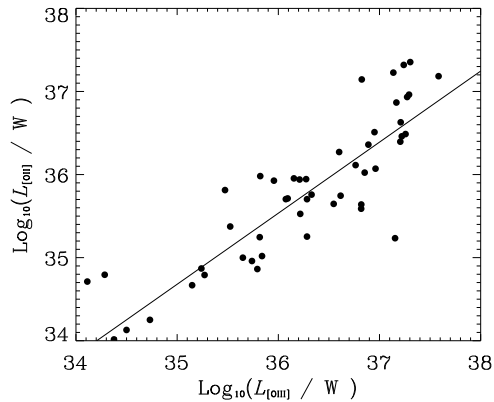}}
\end{picture}
\end{center}
{\caption[junk]{\label{fig:lines32} 
{[OII] emission-line luminosity against [OIII] emission-line luminosity for 
those objects with data in both lines from the 3CRR sample.
The solid line shows the best-fitting relation.}
}}
\end{figure}

In cases where no [OIII] line flux was available, other
narrow-line fluxes were used to estimate it.
Most commonly, the other line flux used was [OII].
There are $57$  objects in the 3CRR sample with both
[OIII] and [OII] measurements. This enabled a best-fitting relation 
to be determined of the form $\log_{10}(L_{{\rm [OII]}}) = a  + 
b \log_{10}(L_{{\rm [OIII]}})$, where $a = 4.74 \pm 1.44, 
b = 0.86 \pm 0.04$ (see Fig.~\ref{fig:lines32}). 
The `best-fit' line was calculated using an algorithm that 
minimises the sum of the squares of the perpendicular distances from the
data points to a line with an adjustable slope and intercept.
One reason why this relation is not a proportionality may be the 
systematic changes of [OII]/[OIII] ratio with narrow-line luminosity
expected if the effective ionization parameter changes as a function of
$L_{\rm phot}$ \citep{sea}: HEGs, which dominate at high $L_{\rm phot}$,
have a systematically lower [OII]/[OIII] ratio than LEGs, which
dominate at low $L_{\rm phot}$.
The most common other lines used
were Lyman-$\alpha$, Mg II, [NeIV], [NeV], CIII], H$\alpha$ and CII]
which were used to estimate [OIII] using the average
line ratios quoted by \citet{mcarthy}.

Fourteen radio galaxies in the 3CRR sample and three radio galaxies in the 
6CE had no emission-line data and 
these objects are identified in Fig.~\ref{fig:pca}. These objects are 
the FRI radio galaxies 3C83.1B, 3C288, 3C296, 3C310, 3C314.1, 3C315, NGC6109, NGC6251,
NGC7385 and 3C465, the FRII LEGs 3C427.1, 6C1143+3703 and 6C1159+3651, and the FRII HEGs 3C68.2, 4C74.16, 3C292 and 6C1129+3710.
Emission-line data for these objects was obtained by generating a random value
with a Gaussian distribution function with the mean given by Eqn.~\ref{eqn:L151Lbol}
and with the observed scatter.

Three 3CRR quasars (3C9, 3C432 and 3C454),
six 6CE quasars (6C0824+3535, 6C0913+3907, 6C1148+3842, 6C1213+3504, 6C1220+3723 and 6C1255+3700)
and twelve 7CRS quasars (5C6.5, 5C6.33, 5C6.34, 5C6.95, 5C6.160, 5C6.237, 5C6.279, 5C6.287, 7C0808+2854, 5C7.70, 5C7.87
and 7C0825+2930) had $\rm{[OIII]}$
emission-line strengths estimated by assuming a rest-frame  equivalent width of $30\rm{\AA}$  \citep*{hjr}, and
the optical frequency power-law index $\alpha_{\rm opt} = -0.44$ for $\lambda < 5000 \rm{\AA}$
and  $\alpha_{\rm opt} = -2.45$ for $\lambda > 5000 \rm{\AA}$
\citep{vandenberk}.
The distribution of quasars with estimates of $L_{\rm [OIII]}$ from equivalent widths is found to be
compatible with the distribution of quasars with real measurements in each of the three samples and
overall.

The line strengths for 6C0955+3844 and 6C1045+3513 were taken from \citet{wsubmm}.
A new spectrum, redshift, classification and $L_{\rm [OII]}$ has been found for 3C325 \citetext{Grimes et al., in prep}.
An upper limit for the $\rm{[OIII]}$ emission-line strength for 3C386 was taken from \citet{s96}.
All objects that had only an upper limit were treated as if the upper limit was the $\rm{[OIII]}$ luminosity, 
but this is not expected to have a significant effect on any result.

A complete updated
list of measured 3CRR line luminosities and their references are available at
http://www-astro.physics.ox.ac.uk/$\sim$cjw/3crr/3crr.html.
A table of the 3CRR, 6CE and 7CRS emission-line strengths,
from either measurements or estimates, is available at
http://www-astro.physics.ox.ac.uk/$\sim$sr/grimes.html.

\subsection{Radio source counts}
\label{sec:scdata}
The number of known radio sources at $151$ MHz is very much larger than the
sources in the complete samples with redshifts described above and is a valuable extra
constraint. The source counts 
have been determined
from the 6C \citep{hbw} and
7C \citep{mcg} surveys, which have much larger sky areas than the 6CE and 7CRS redshift surveys.
 The 7C source counts go as faint
as $0.1$ Jy, and so provide a low-flux-density constraint on the
GLF, even in the absence of redshift information.

\section{Principal Components Analysis}
\label{sec:pca}
PCA is a statistical technique that has been extensively used in astrophysics
 for the analysis of spectral data
(e.g. \citealt{folkes, madge}) and for reducing the dimensionality of problems.
It finds how a set of properties of a sample of objects are inter-related, 
by identifying sets of parameters that always correlate and are most discriminatory
between each source in the sample. The strength of each new parameter is measured
by the amount it contributes to the variance of the sample. Sub-samples of data
are identified from any clustering in the space defined by the new axes of the PCA.

There have been many treatments of PCA in the literature (e.g. \citealt{fw, mh}).
PCA searches for the best-fitting set of axes to replace the initial set of axes corresponding
to the attributes of the data, using the squared deviation of the points from the axes as
the goodness-of-fit criterion and by enforcing orthogonality. This process gives rise to an eigenvalue equation.
The matrix formed from all the eigenvectors rotates the original basis to a new set of orthogonal axes. 

Practically, it is necessary to standardize the variables under analysis, so that the components
of the normalized attributes $X = \{x_{ij}\}$ are given by

\begin{equation}
\label{eqn:norm}
x_{ij} = \frac{r_{ij} - \bar{r}_j}{s_j \sqrt{n}},
\end{equation}

\noindent where the index $i$ runs from $1$ to $n=302$, the total number of objects in the 3CRR, 6CE and 7CRS samples, 
and $j = 1, 2$, so that $r_1 = \log_{10}L_{151}, r_2 = \log_{10}L_{\rm [OIII]}$, for each attribute 
$\bar{r}$ is the mean and $s$ the standard deviation.

\subsection{Results of the PCA}

\begin{table}
\begin{center}
\begin{tabular}{lcr}
\hline\hline
&\multicolumn{1}{c}{$\lambda_{\alpha}$}&\multicolumn{1}{c}{$\lambda_{\beta}$}\\
\hline
Eigenvalue&$1.86$&$0.14$ \\
Proportion&$0.93$&$0.07$ \\
\hline
&\multicolumn{1}{c}{$\mathbf{e}_{\alpha}$}&\multicolumn{1}{c}{$\mathbf{e}_{\beta}$}\\
\hline
$\log_{10}L_{151}$ normalized&$\frac{1}{\sqrt{2}}$&$\frac{1}{\sqrt{2}}$\\
\\
$\log_{10}L_{\rm [OIII]}$ normalized&$\frac{1}{\sqrt{2}}$&$-\frac{1}{\sqrt{2}}$\\
\hline\hline
\end{tabular}
\caption[PCA table]{\label{tab:pca}{The eigenvalues $\lambda_{\alpha}$ and $\lambda_{\beta}$ and eigenvectors $\mathbf{e}_{\alpha}$ and $\mathbf{e}_{\beta}$ in terms of the normalized components of $\log_{10}L_{151}$ and $\log_{10}L_{\rm OIII}$ for the principal components analysis.}} 
\end{center}
\end{table}

The PCA was performed and the average value of $\log_{10}L_{151}$ was found to be $26.73$ with a spread of
$1.05$ and the average value of $\log_{10}L_{\rm [OIII]}$ was $35.46$ with a spread of $1.10$.
This gave rise to the eigenvalues and eigenvectors given in Table \ref{tab:pca}.
As can be seen, the first eigenvector contributes $93$ per cent of the total scatter.
Also, the principal eigenvector $\mathbf{\alpha}$ shows that the dominant relation
between $\log_{10}L_{151}$ and $\log_{10}L_{\rm [OIII]}$ is that they are positively correlated,
(with an equal contribution from their \textit{normalized} components because there are only
two components), and the secondary
relation is an anti-correlation, much smaller in magnitude than the main correlation.

We have now defined two new axes on which to define a GLF.
The projections of the attributes on these new axes are given by $\alpha$
and $\beta$, where 
\begin{equation}
\label{eqn:alpha}
\alpha_i = \frac{1}{\sqrt{2}} x_{i1} + \frac{1}{\sqrt{2}} x_{i2},
\end{equation}
and similarly 
\begin{equation}
\label{eqn:beta}
\beta_i = \frac{1}{\sqrt{2}} x_{i1} - \frac{1}{\sqrt{2}} x_{i2}.
\end{equation}

\begin{figure}
\begin{center}
\setlength{\unitlength}{1mm}
\vspace{1.7cm}
\begin{picture}(100,65)
\put(5,10){\includegraphics{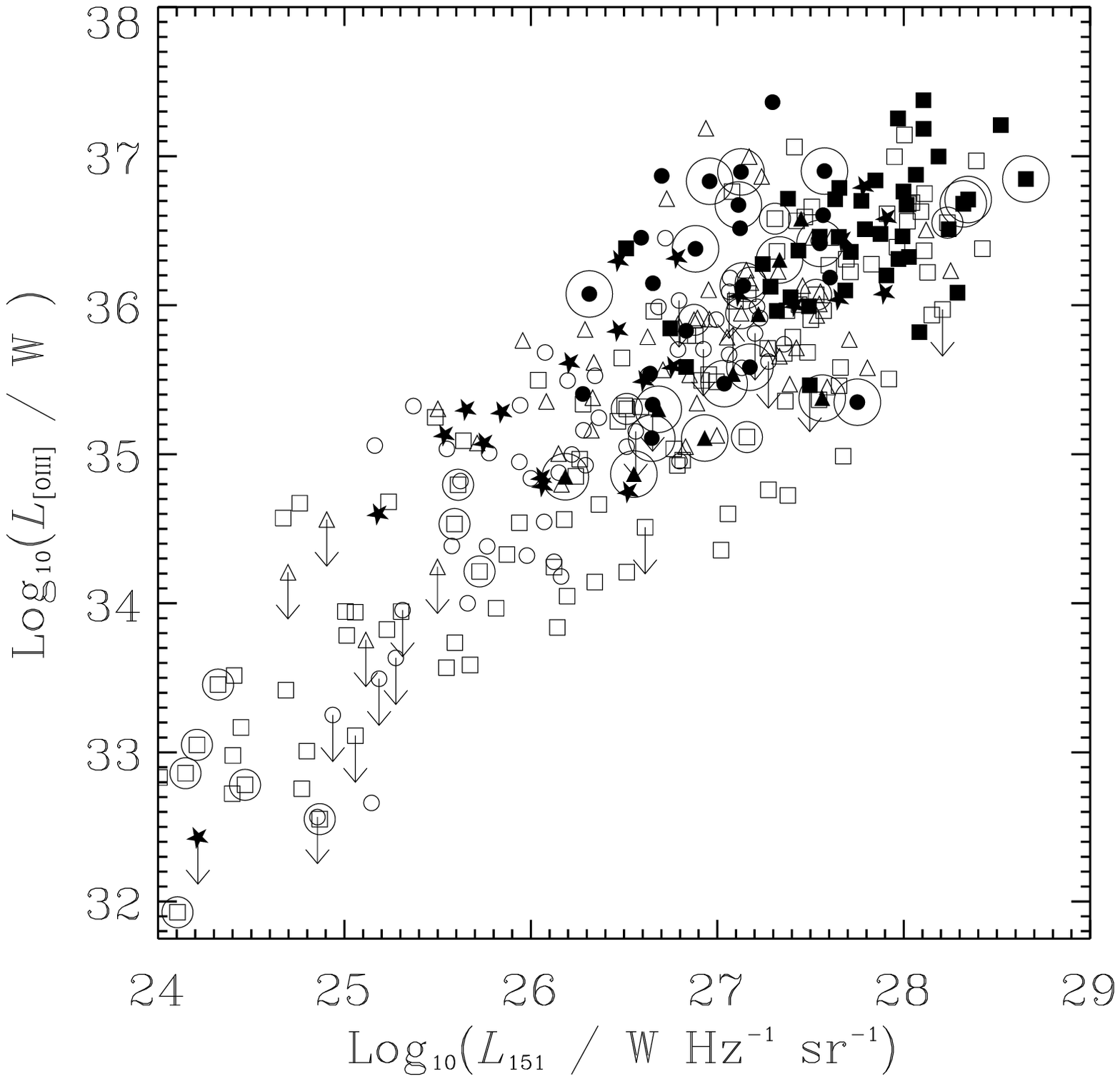}}
\end{picture}
\vspace{0.3cm}
\begin{picture}(100,65)
\put(5,10){\includegraphics{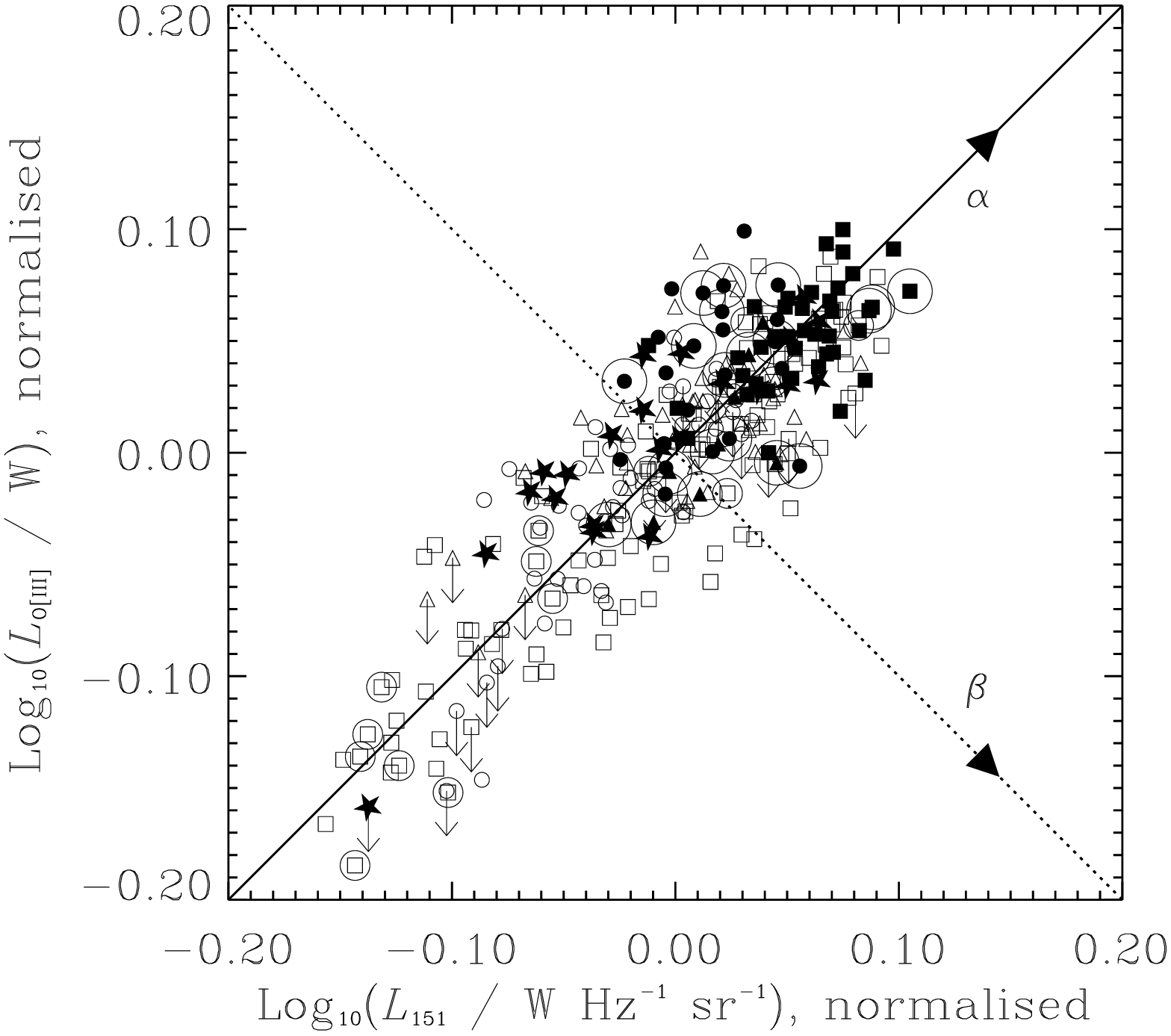}}
\end{picture}
\begin{picture}(100,65)
\put(5,10){\includegraphics{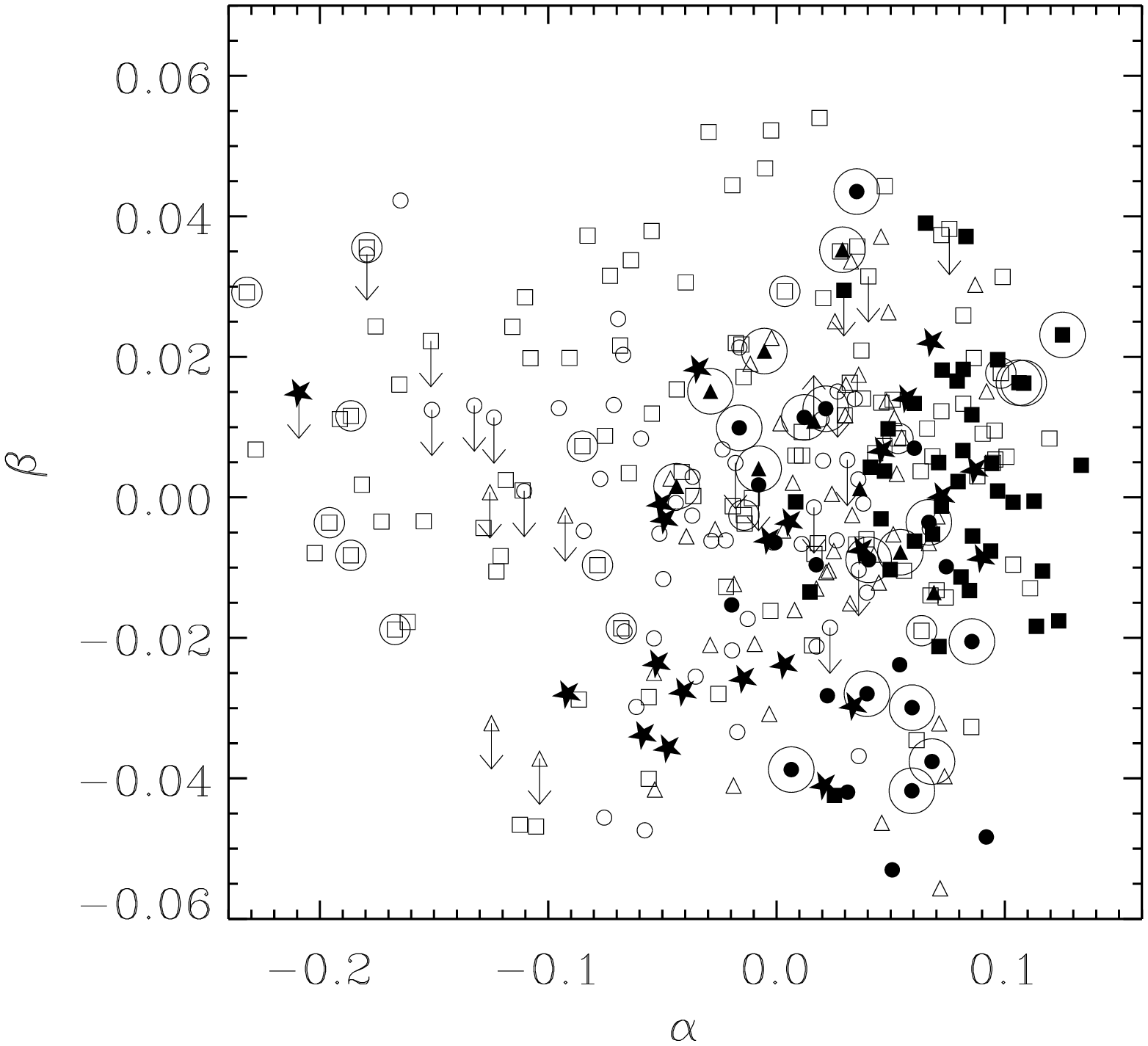}}
\end{picture}
\end{center}
\vspace{-2.5cm}
{\caption[junk]{\label{fig:pca} 
{Top: The $\log_{10}L_{\rm [OIII]} - \log_{10}L_{151}$ plane for
the 3CRR, 6CE and 7CRS radio galaxies and quasars. The symbols are
as in Fig.~\ref{fig:pz} for all figures.
Large circles surrounding open symbols identify radio galaxies that
have estimates of $L_{\rm [OIII]}$ from the $\log_{10}L_{\rm [OIII]} - \log_{10}L_{151}$
correlation with the observed scatter. Larger circles surrounding filled symbols identify
quasars that have $L_{\rm [OIII]}$ estimated by assuming a value of $30\rm{\AA}$ 
for the
rest-frame equivalent width of the $\rm{[OIII]}$ line.
Centre: The normalized $\log_{10}L_{{\rm [OIII]}} - $
normalized $\log_{10}L_{151}$ plane. 
The eigenvectors corresponding to the first principal component (solid line) and
second principal component (dotted line) are also shown.
Directions of increasing $\alpha$ and $\beta$ are marked.
Bottom: The distribution of radio quasars and radio quasars in the 
$\alpha-\beta$ plane. 
}}}
\end{figure}

From Fig.~\ref{fig:pca}
we can see that the first transformation of the data (top
to the centre plot) involves a normalization (Eqn.~\ref{eqn:norm}) so that the axes are expanded. 
The second transformation of the data (centre to bottom plot) is a rotation (Eqns.~\ref{eqn:alpha} and ~\ref{eqn:beta})
of the data.
There seems to be a cutoff in $\alpha$-space at $\sim -0.10$, so
that below this value there is only one (weak) quasar. Also, the radio quasars 
are biased to lower values of $\beta$, because they have higher values of
$L_{\rm [OIII]}$ for a given value of $L_{151}$.

Notice that the objects from the 3CRR, 6CE and 7CRS samples can all be
described by the same principal components. Since it can be seen that
for any flux-density-limited sample, $L_{151}$ and redshift are correlated
(see Fig.~\ref{fig:pz}), it might be expected that different samples
may have different principal components. Tests have shown that these 
components do not vary significantly for various combinations of samples.
The most obvious conclusion is that the correlation between emission-line
and radio luminosities is much more important than any $L_{151}- z$ 
correlations \citep{wem}. It can be shown from the normalized components that the relationship
between $L_{[\rm OIII]}$ and $L_{151}$ can be expressed as
 
\begin{equation}
\label{eqn:L151Lbol}
\left(\frac{L_{\rm [OIII]}}{\rm W}\right) = 3.4 \times 10^7 \times \left( \frac{L_{151}}{\rm W\,Hz^{-1}\,sr^{-1}}\right)^{1.045}
\end{equation}

For comparison, \citet{ser} found that there was a radio-optical correlation in 
steep-spectrum quasars of the form $L_{\rm phot} \propto L_{408}^{0.6 \pm 0.1}$.
\citet{wqrlf} found a less steep slope of $L_{\rm phot} \propto 
L_{151}^{0.4 \pm 0.1}$ for $24$ radio loud quasars from
7CRS, and attributed the discrepancy at least
in part to the fact that
\citet{ser} did not impose a limiting absolute magnitude
in their definition of a quasar. 
\citet{wconf} found a slope of $1.00 \pm 0.04$ in the
$\log_{10}L_{\rm [OII]} - \log_{10}L_{151}$ relation.
All these previous studies adopted $\Omega_{\rm M} = 1.0, \Omega_{\Lambda} = 0.0$.
The differences between these slopes can probably largely be attributed to 
the subtle differences between quasars and radio galaxies. 

Values of $\alpha$ and $\beta$ for all objects in the samples are available from
the table at http://www-astro.physics.ox.ac.uk/$\sim$sr/grimes.html.

\section{Generalized luminosity function}
\label{sec:model}
In this section, we will define generalized luminosity functions (GLFs) assuming that all of the radio sources are
drawn from the same population, for comparison with two-population GLFs, which
will be presented in Sec.~\ref{sec:2pop}. A GLF $\rho(\alpha, \beta, z)$  
is modelled as

\begin{equation}
\label{eqn:lf}
\rho(\alpha, \beta, z) = \rho_0 10^{-p(\alpha - \alpha_{\rm break})} g(\beta) f(z),
\end{equation}

\noindent where
\begin{equation}
f(z) =  \exp\left(- \frac{(z - z_0)^2}{2z_1^2}\right),
\end{equation}

\noindent and
\begin{equation}
\label{eqn:gbeta}
g(\beta) =  \exp\left(- \frac{(\beta-\beta_o)^2}{2 \sigma_{\beta}^2}\right),
\end{equation}

\noindent and $\alpha$ is the first principal component, $\beta$ is the second 
principal component, $\sigma_{\beta}$ is the scatter, $\rho_0$ is the 
normalization and  $z_1$ and $z_0$ define the Gaussian evolution in redshift
$f(z)$ which fits the RLF in \cite{wrlf}.
The GLF is defined as the number of sources per unit $\alpha$, per unit $\beta$
per unit volume.
Gaussian scatter in a direction perpendicular to the best-fitting slope of
$\log_{10}L_{\rm [OII]} - \log_{10}L_{151}$ is seen by \citet{wconf},
so Gaussian scatter in $\beta$ is assumed.
The function is a broken power-law in $\alpha$, so that
$p = p_1$ when $\alpha > \alpha_{\rm break}$ and $p = 0$ otherwise.
The $\beta$-offset parameter $\beta_o$ accounts for the fact that 
a GLF $\rho(\alpha, \beta, z)$ derived from a symmetric distribution of
objects in the $(\alpha, \beta)$-plane will produce a distribution that has
an offset in the $\beta$-direction for a flux-density limited sample. 
This arises because, for a given redshift, we select only the highest
values of $L_{151}$. From the definition of $\alpha$, this means
that we see only objects with low values of $L_{\rm [OIII]}$ for a given value
of $\alpha$, and thus we see objects with less negative values of $\beta$
than would be seen in a sample that was not flux-density-limited.
An analogous problem is that the distribution of sources with 
respect to spectral index in a complete radio-selected sample varies with the
frequency at which the sample is selected, \citep{kell, wb}.

We consider both a simple unified scheme with constant $\Theta_{{\rm trans}}$ and a receding-torus scheme with
\begin{equation}
\label{eqn:thetatrans}
\tan \Theta_{{\rm trans}} = (L_{{\rm [OIII]}}/L_0)^{\frac{1}{2}} \tan \Theta_0, 
\end{equation}
\noindent where $L_0$ is taken as the median value of $L_{\rm [OIII]}$.

A factor of $\sin\theta$ accounts for the probability of detecting an object inclined at an angle $\theta$ to the 
line of sight. The detected object is a quasar if 
$\theta < \Theta_{\rm trans}$ and a radio galaxy 
if $\theta \ge \Theta_{\rm trans}$. Therefore the GLF for radio galaxies
 is given by

\begin{equation}
\rho_{\rm RG}(\alpha, \beta, z) = \rho(\alpha, \beta, z) \times \cos\Theta_{{\rm trans}}, 
\end{equation}
and the GLF for radio quasars is given by
\begin{equation}
\rho_{\rm Q}(\alpha, \beta, z) = \rho(\alpha, \beta, z) \times (1 - \cos\Theta_{{\rm trans}}).
\end{equation}
We will denote the GLF with the receding torus as 1R and with the simple
unified scheme as 1S.

A maximum-likelihood analysis was
performed to optimize the free parameters of the GLF. For both GLFs, 
there are eight free parameters: $\sigma_{\beta}, \rho_0, p_1, \alpha_{\rm break}, \beta_o, z_0, z_1$ and $\Theta_0$.
In order to maximise the likelihood we must 
minimize $S$, where $S = -2 \ln (\mbox{likelihood})$. Following
\citet{marsh}, we find an expression for $S$,

\begin{eqnarray}
\label{eqn:maxlike}
S&=& -2\sum_{i=1}^{\rm NQ} \ln[\rho(\alpha_i, \beta_i, z_i) (1 - \cos\Theta_{{\rm trans}})] \\
\nonumber
&-&2\sum_{i=1}^{\rm NRG} \ln[\rho(\alpha_i, \beta_i, z_i) \cos\Theta_{{\rm trans}}]  \\
\nonumber
&+&2\int\!\!\!\int\!\!\!\int \rho_{\rm Q} \Omega \frac{{\rm d}V}{{\rm d}z} \left| J \right| \rm{d}z \rm{d}\log_{10}L_{151} \rm{d}\log_{10}L_{\rm [OIII]}\\
\nonumber
&+&2\int\!\!\!\int\!\!\!\int\,\rho_{\rm RG}\Omega \frac{{\rm d}V}{{\rm d}z} \left| J \right|  \rm{d}z \rm{d}\log_{10}L_{151} \rm{d}\log_{10}L_{\rm [OIII]},\\
\nonumber
\end{eqnarray}

\noindent where NQ is the number of radio quasars including weak quasars, NRG is the number
of radio galaxies, $\Omega = \Omega(\log_{10}L_{151}, z)$ is the sky
area available from the samples for this value
of $z$ and $L_{151}$, $J$ is the Jacobian matrix for the change of variables from principal components
to $\log_{10}L_{\rm [OIII]}$ and $\log_{10}L_{151}$, and ${\rm d}z \times
{\rm d}V/{\rm d}z$ is the differential co-moving volume element. 
In essence, there are two terms for quasars, and two for radio galaxies: 
one is the sum over all NQ quasars
or NRG radio galaxies in the samples,  and the last two terms are the integrals over the model being tested and
should give $\approx 2$ NQ and $\approx 2$ NRG respectively for good fits.

The errors associated with the parameters were found by 
calculating the components of the Hessian matrix ($\nabla \nabla S$)
at the location of the minimum, inverting this matrix to obtain the
covariance matrix ($[\sigma^2]_{ij} = 2[(\nabla \nabla S)^{-1}]_{ij}$),
and taking the $1\sigma$ errors as given by the square root of the diagonal
elements of this matrix (e.g. \citealt{sivia}).

To find relative probabilities for the GLFs with or without a receding-torus,
the procedure set out by \citet{sivia} was used. 
The ratio of the posterior probabilities of GLF 1R with respect to GLF
1S, $P_{\rm RS}$,  is given approximately by

\begin{eqnarray}
\nonumber
\label{eqn:rel}
P_{\rm RS} &=& \frac{P(\rm 1R|data)}{P(\rm 1S|data)} \\
 &=& \frac{ e^{\frac{-S_{\rm 1R | min}}{2}} \sqrt{\det(\nabla \nabla S_{\rm 1S} |_{S_{\rm min}})} }{ e^{\frac{-S_{\rm 1S | min}}{2}} \sqrt{\det(\nabla \nabla S_{\rm 1R} |_{S_{\rm min}})} } \times F,
\end{eqnarray}

\noindent where $\det (\nabla \nabla S_{\rm 1R} |_{S_{\rm min}})$ is the determinant of the 
Hessian matrix for GLF 1R, evaluated at $S = S_{\rm min}$. All of the free 
parameters are common to both GLFs, except for $\tan \Theta_{\rm trans}$ which 
is coded in a different way in each GLF (see Equation \ref{eqn:thetatrans}) but the 
prior ranges on this parameter are equal, giving $F = 1$.

\subsection{Results}
\label{sec:1popresults}

The maximum-likelihood routine was run for both the GLF with a receding torus
1R and the GLF with the simple constant-$\Theta_{{\rm trans}}$ unified scheme 1S. 
A downhill simplex \citep{press} method was used to locate the best-fitting 
parameters, and these are presented in Table \ref{tab:param1}. 
GLF 1R was found to be more likely than GLF 1S by a factor of $\sim 10^{7}$.
The difference in likelihood between 1R and 1S arises mainly from the differences
in their unified scheme parameters, as most of the parameters describing the
shape of the GLF are very similar.
This naively implies that the simple non-luminosity-dependent unified scheme is strongly ruled out,
as there exists an alternative physical model which is far more probable
given the data. This has put on a quantitative basis the obvious
(e.g. \citealt{law}) statement that a constant-quasar-fraction 
model is inconsistent with the data.
In Fig.~\ref{fig:pca},
it is clear that the quasar fraction changes dramatically with $\alpha$.

\begin{table*}
\begin{center}
\begin{tabular}{ccccccccc}
\hline\hline
&$\log(\rho_0)$&$p_1$&$\alpha_{\rm break}$&$z_0$&$z_1$&$\log_{10}(\sigma_{\beta}^2)$&$\beta_o$&$\Theta_0$\\
\hline
1R & $-2.049 \pm 0.031$ & $27.08 \pm 1.35$ & $-0.058 \pm 0.005$ & $1.951 \pm 0.057$& $0.550 \pm 0.032$
  & $-3.340 \pm 0.048$ &$-0.021 \pm 0.003$&$42.0^{+3.9}_{-3.9}$ \\
\hline
1S &$-2.060 \pm 0.097$&$27.06 \pm 1.73$ & $-0.058 \pm 0.006$&$1.940 \pm 0.058$&$0.545\pm 0.039$
  & $-3.342 \pm 0.051$ &$-0.020\pm0.003$& $45.8^{+3.0}_{-3.0} $ \\
\hline\hline
\end{tabular}\\
{\caption[Table of parameters]{\label{tab:param1}{Best-fit parameters
for the Model 1R and Model 1S.
For the simple unified scheme, $\Theta_0$ is the constant transition angle, $\Theta_{\rm trans}$, whereas 
for the receding torus model, $\Theta_0$ is related to $\Theta_{\rm trans}$ by Eqn.~\ref{eqn:thetatrans},
using the median $\log_{10}(L_0) = 35.405$ as a reference. The values of $S_{\rm min}$ are $8701.89$
for 1R and $8734.37$ for 1S. As the values of $\det(\nabla \nabla S)$ for the 
two models are comparable, 1R is more likely than 1S by $\sim 10^7$.
}}}
\end{center}
\end{table*}

\citet{wrlf} derived three models for the RLF at $151$ MHz
from the low-frequency-selected 3CRR, 6CE and 7CRS samples. 
The GLF in 
$\alpha$ and $\beta$ space, $\rho(\alpha, \beta, z)$ can be 
easily transformed to give a RLF $\rho(\log_{10}L_{151}, z)$ for comparison

\begin{eqnarray}
\rho(\log_{10}L_{151}, z) = \int \rho(\alpha, \beta, z) \left| J \right| \frac{\rm{d}V}{\rm{d}z} \rm{d}\log_{10}L_{{\rm [OIII]}}.
\end{eqnarray}

\begin{figure}
\begin{center}
\setlength{\unitlength}{1mm}
\begin{picture}(150,60)
\put(0,-10){\includegraphics{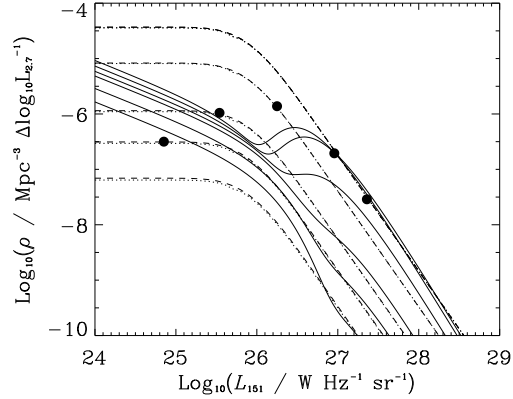}}
\end{picture}
\end{center}
{\caption[junk]{\label{fig:lf} 
{The radio luminosity function at $151$ MHz and $z = 0.0001, 0.25, 0.5, 1.0, 2.0, 3.0$ from \citet{wrlf} (solid lines)
compared with the RLF from the optimized GLFs 1R (dashed lines) and 1S (dotted lines). 
The circles
show the location of the lower limit to the flux density of
the 7CRS sample for each redshift, as indicated on Fig.~\ref{fig:pz}
which shows the luminosity at each
redshift below which the GLF is not well constrained.
The $z=2$ and $z=3$ curves are identical, but the limit for
$z=2$ is above that for $z=3$.}}}
\end{figure}

Fig.~\ref{fig:lf}
shows the Model B RLF from \citet{wrlf}, which has been adapted to a 
$\Omega_{M} = 0.3, \Omega_{\Lambda} = 0.7, H_0 = 70$ km s$^{-1}$ Mpc$^{-1}$ cosmology,
and the RLFs derived from the GLFs 1R and 1S, which have almost exactly the same shape.
Note that agreement with \citet{wrlf} is only expected to be good above points which represent
the flux-density limit of 7CRS. This is because only $L_{151}-z$ plane data
has been used to constrain the GLF, but the \citet{wrlf} RLF is also constrained
by source counts.

\begin{figure}
\begin{center}
\setlength{\unitlength}{1mm}
\begin{picture}(150,60)
\put(0,-10){\includegraphics{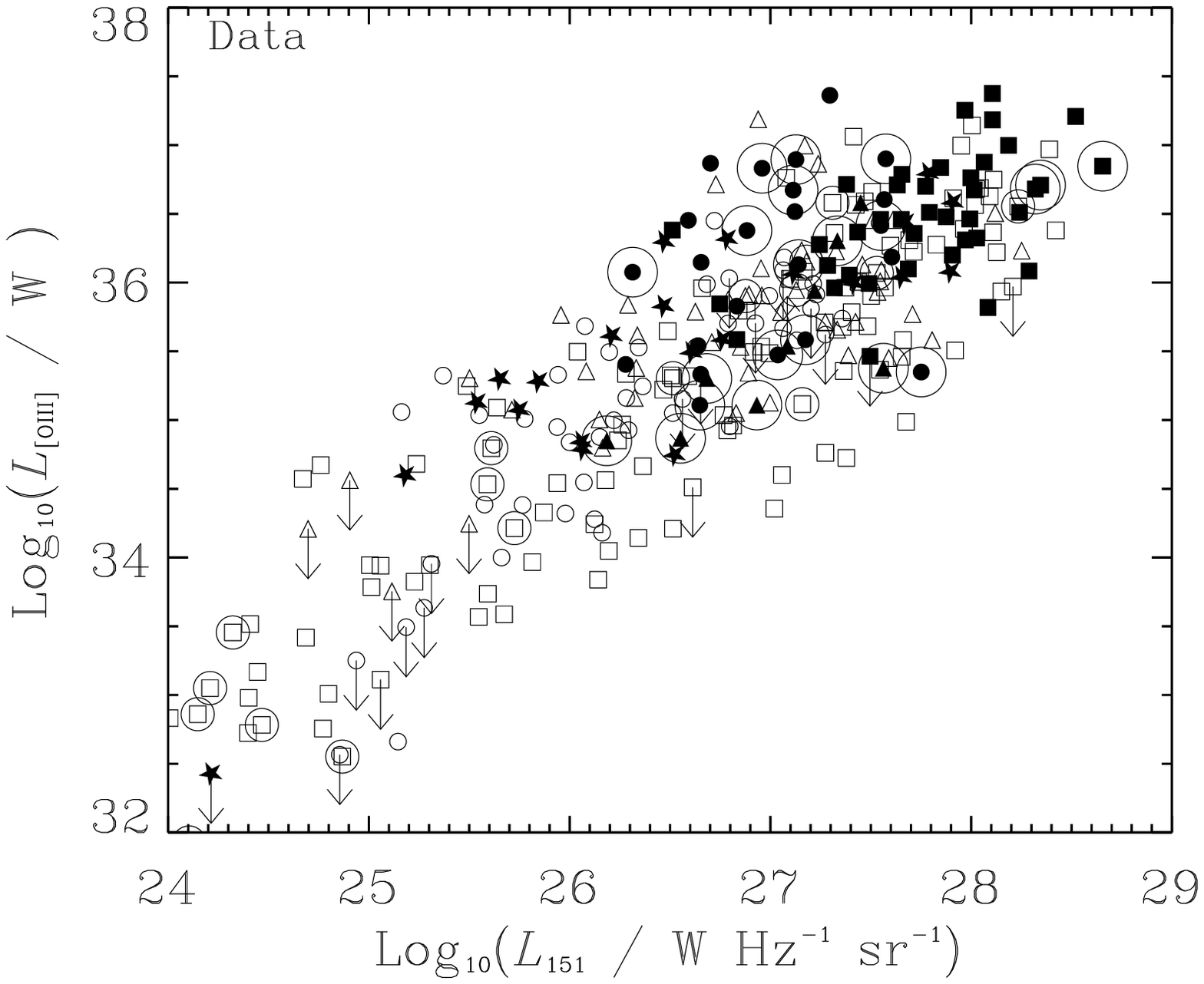}}
\end{picture}
\begin{picture}(150,60)
\put(0,-10){\includegraphics{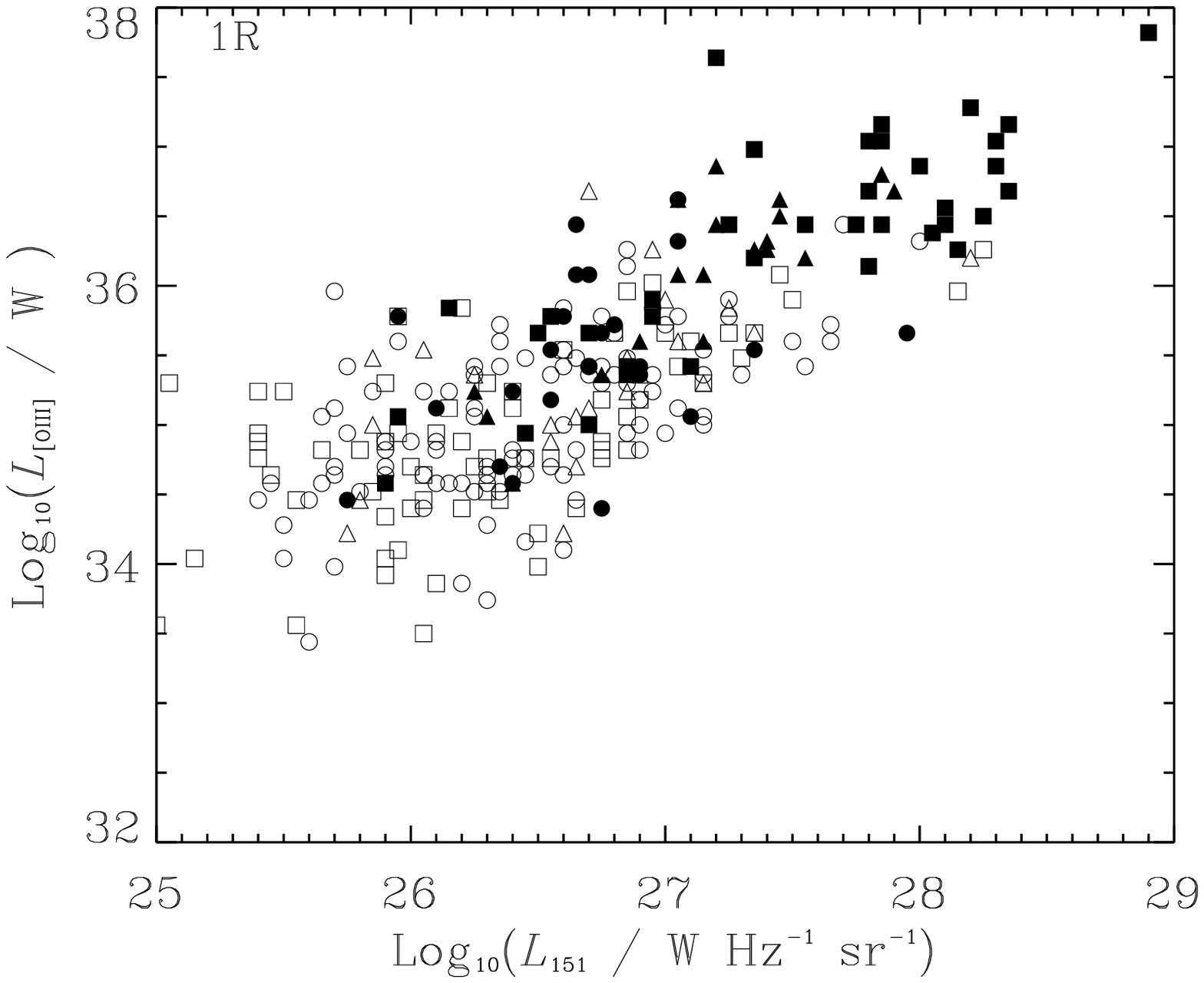}}
\end{picture}
\begin{picture}(150,60)
\put(0,-10){\includegraphics{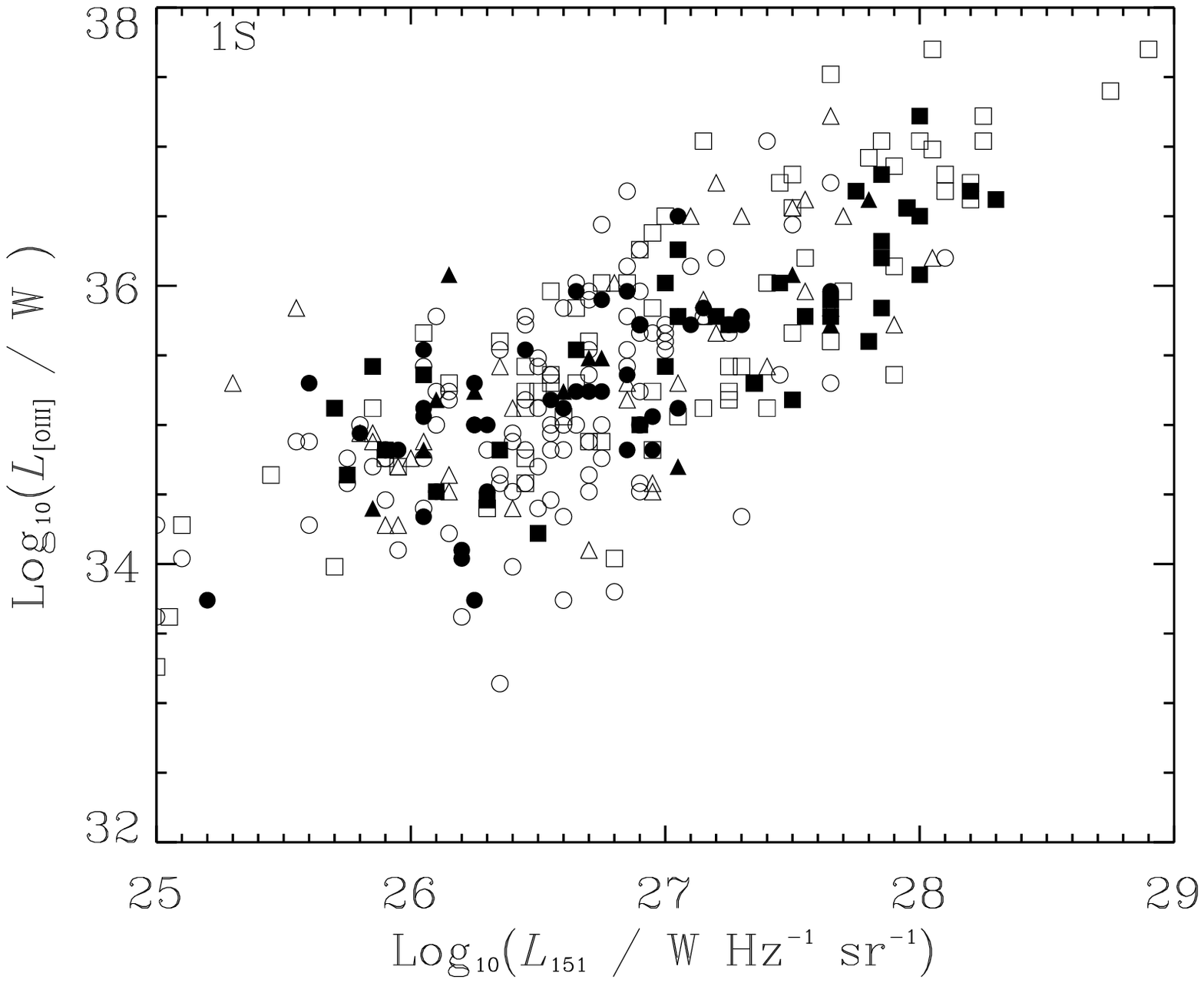}}
\end{picture}
\end{center}
{\caption[junk]{\label{fig:mc} 
{The $\log_{10}L_{\rm [OIII]} - \log_{10}L_{151}$ plane for the 3CRR, 6CE and 7CRS radio galaxies and radio quasars (top) and simulations of $\log_{10}L_{\rm [OIII]}$ and  $\log_{10}L_{151}$ data from 
the 1R (centre) and 1S (bottom) GLFs. 
Symbols are as in Fig.~\ref{fig:pz}.
}}}
\end{figure}

Simulations of the $\log_{10}L_{\rm [OIII]} - \log_{10}L_{151}$
plane from GLFs 1R and 1S are shown in Fig.~\ref{fig:mc}, and show that these
simple models can reproduce the data reasonably well 
at high values of $L_{151}$ and $L_{\rm [OIII]}$. One of the key features of the data, namely a lack of
quasars at low values of $L_{\rm [OIII]}$, is seen in the simulations from GLF 1R, 
whereas the simulations from 1S clearly do not cause a difference in emission-line
luminosities between radio quasars and radio galaxies. It is therefore easy to see why there
is such a huge difference in likelihoods.

\begin{figure}
\begin{center}
\setlength{\unitlength}{1mm}
\begin{picture}(150,60)
\put(0,-10){\includegraphics{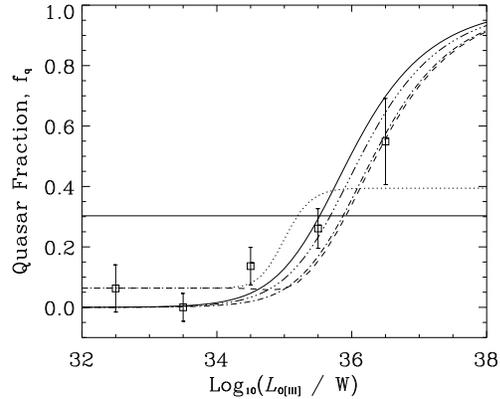}}
\end{picture}
\end{center}
{\caption[junk]{\label{fig:fq}{The quasar fraction against [OIII]
luminosity for GLFs 1R (solid line), 1S (horizontal solid line), 2RS (dashed line), 2SS (dotted line) and 2RR (dot-dashed line).
Also shown is the receding torus model from \cite{wqf} converted to the correct cosmology and using [OIII] instead of [OII], which was originally used (triple-dot-dashed line). 
The squares show the quasar fractions for the 3CRR, 6CE and 7CRS datasets in
bins of $\log_{10}L_{\rm [OIII]}$ luminosity.
The error bars show the $\sqrt N$ errors.}}}
\end{figure}

\begin{figure}
\begin{center}
\setlength{\unitlength}{1mm}
\begin{picture}(150,60)
\put(0,-10){\includegraphics{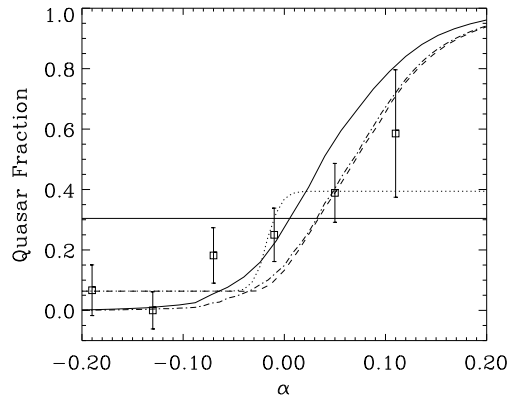}}
\end{picture}
\end{center}
{\caption[junk]{\label{fig:qfalpha}
{The quasar fraction against $\alpha$, for GLFs 1R (solid line), 1S (horizontal solid line),
2RS (dashed line), 2SS (dotted line) and  2RR (dot-dashed line). The squares
show the quasar fractions for the 
3CRR, 6CE and 7CRS datasets in bins of $\alpha$. 
The error bars show the $\sqrt N$ errors. }}}
\end{figure}

We can compare the receding-torus parameter with \citet{wqf}
and \cite{simp}. Fig.~\ref{fig:fq} shows the quasar fraction as a function
of $L_{\rm [OIII]}$ for GLFs 1R and 1S, as well as two-population GLFs which will be
introduced in Sec.~\ref{sec:2pop}. It is clear that a constant-transition-angle unified
scheme cannot
reproduce the quasar fraction data from the 3CRR, 6CE and 7CRS samples
across the range of $L_{\rm [OIII]}$, while the one-population GLF with 
a receding torus  
provides a reasonably good fit to the data. Note that the quasar fraction
curve derived in \citet{wqf} does not include FRI objects.
The same conclusions can
be drawn from Fig.~\ref{fig:qfalpha}, where the quasar fraction is
plotted as a function of $\alpha$. The quasar fraction from GLF 1R also agrees reasonably well
with the receding-torus type model parameters derived from 3CRR and 7CRS
$L_{\rm [OII]}$ data from \citet{wqf}.

The scatter in the $\beta$-direction, $\sigma_{\beta}$ can be 
related to the scatter in the $L_{151}-L_{\rm [OIII]}$ relation 
by the factor $s_2 \sqrt{2 n}$, giving a value of $0.58$, which
is similar to the scatter of $\sim 0.6$ in the radio-optical correlation found by \cite{ser},
and the scatter in the $L_{\rm[OII]} - L_{151}$ correlation of $0.54$ found by \cite{wconf}.

\section{A two-population model}
\label{sec:2pop}
The results of Sec.~\ref{sec:1popresults} imply that some mechanism is needed to
cause the change in quasar fraction with $\alpha$, recalling that $\alpha$
is a measure of the strongly correlated radio and $[{\rm OIII}]$ luminosities. 
An alternative mechanism to the receding-torus
scheme could be a two-population model composed of a high-$\alpha$ population of radio
galaxies and radio quasars related by a unified scheme, in addition to a low-$\alpha$
population mostly composed of radio galaxies, but with a unified scheme that allows
a small fraction of quasars, to account for objects such as 3C386.
This would certainly cause an increase in quasar fraction with $\alpha$.

A two-population GLF was created by adding together a low-$\alpha$ population
GLF ($\rho_1$) to a high-$\alpha$ population GLF ($\rho_2$). This method is
similar to that of \citet{wrlf} for the construction of the $151$
MHz RLF, and utilises the same evolutionary 
forms of both populations. Thus,

\begin{table}
\begin{center}
\begin{tabular}{lll}
\hline
\hline
GLF&One-population schemes&\\
\hline
1S&Simple&\\
1R&Receding torus&\\
\hline
&\multicolumn{2}{c}{Two-population Schemes}\\
GLF&Population 1&Population 2\\
\hline
2SS&Simple&Simple\\
2RS&Simple&Receding torus\\
2RR&Receding torus&Receding torus\\
\hline
\hline
\end{tabular}
\end{center}
\caption{Summary of the unified schemes used in each population for each GLF.}
\end{table}

\begin{equation}
\label{eqn:glf1}
\rho_1 = \rho_{10} 10^{-p_1(\alpha - \alpha_{\rm cut,1})}  
\exp \left[{- 10^{q(\alpha - \alpha_{\rm cut,1})} }\right] g(\beta) f_1(z),
\end{equation}

\noindent where $g(\beta)$ is defined in Eqn.~(\ref{eqn:gbeta}) and \\

\begin{tabular}{lclcl}
$f_1(z)$&$=$&$(1 + z)^k$&for&$z<z_1$ \\
&$=$&$(1 + z_1)^k$&for&$z \geq z_1.$ \\
\end{tabular}

\noindent Also

\begin{equation}
\label{eqn:glf2}
\rho_2 = \rho_{20} 10^{-p_2(\alpha - \alpha_{\rm cut,2})} 
\exp \left[{- 10^{q(\alpha - \alpha_{\rm cut,2})} }\right] g(\beta) f_2(z),
\end{equation}

\noindent where \\

\begin{tabular}{lclcl}
$f_2(z)$ &$=$&$ \exp\left[{ - \frac{1}{2} (\frac{z - z_{2\rm{a}}}{z_{2\rm{b}}})^2}\right]$&for&$z<z_{2\rm{a}}$\\
      &$=$&$1$&for&$z \geq z_{2\rm{a}}$. \\
\end{tabular}

There is an exponential fall-off in the density of population $1$ 
for $\alpha > \alpha_{cut}$ and in the density of 
population $2$ objects when $\alpha < \alpha_{cut}$. A factor 
$q = s_1 \sqrt n / \sqrt 2$ is the factor by which the range in
$\alpha$ is smaller than the range in $\log_{10} L_{151}$, and acts to make
the exponential decline happen as in \cite{wrlf}. 
Fig.~\ref{fig:rhoalp} 
shows the two-population GLF.

The first GLF we consider is a two-population scheme where the 
high-$\alpha$ population $2$ consists of radio quasars and radio
galaxies unified by a receding-torus
scheme, and the low-$\alpha$ population $1$ consists of radio galaxies and
quasars with a simple unified scheme. Henceforth we will refer to this GLF as 2RS. We also consider the
case where the both the low- and high-$\alpha$ populations are unified by a simple unified scheme and
we will refer to this as the 2SS GLF. In addition, we consider a GLF for a two-population scheme which has quasars and
radio galaxies unified by a receding torus model in both populations, which
will be referred to as Model 2RR.
The radio galaxy
and quasar populations densities are given by

\begin{eqnarray}
\rho_{\rm RG}(\alpha, \beta, z) &=& \rho_1 \cos \Theta_{\rm trans_1} + \rho_2 \cos \Theta_{\rm trans_2}. \\
\nonumber
\rho_{\rm Q}(\alpha, \beta, z)  &=& \rho_1(1 - \cos \Theta_{\rm trans_1}) + \rho_2(1 - \cos \Theta_{\rm trans_2}).
\end{eqnarray}

These GLFs are constrained by the 3CRR, 6CE and 7CRS datasets as before 
(Sec.~\ref{sec:1popresults}) and, because of the increased complexity of 
the two-population GLFs, we need the additional constraint of the 6C and 7C
source counts (described in Sec.~\ref{sec:scdata}).
In addition, the \citet{wrlf} RLF had the constraint of the local
RLF. Here we use the more recently determined local ($z \simeq 0.25$) RLF for AGN from the 2dF
Galaxy Redshift Survey \citep{sad}. We simply use their normalization
of the local RLF at $\log_{10}(L_{151} / {\rm W\,Hz^{-1}\,sr^{-1}}) = 24.0$
(converted from $1.4$ GHz to $151$ MHz using $\alpha_{\rm rad} = -0.8$)
as a prior. Without the use of this we
would have very little constraint on the faint end of the GLF at low
values of redshift. The slope $p_1$ of the  
low-$\alpha$ population can similarly be fixed at a value $6.4$, derived
using the results of \citet{sad}.

The function to be minimised is now defined as
\begin{equation}
S_{\rm total} = A\ln(\chi^2_{\rm A}) + B\ln(\chi^2_{\rm B}) + C\ln(\chi^2_{\rm C}),
\end{equation}
where
the parameters $A, B$ and $C$ are `Lagrange multipliers' or Bayesian
`hyper-parameters' \citep{lahav}, which effectively control the
relative weight applied to the data sets. The function to be minimised
can be shown to be 

\begin{equation}
\label{eqn:hyper}
S_{\rm total} = N_{\rm A} \ln(S) + N_{\rm B} \ln(\chi^2_{\rm sc}) + N_{\rm C}\ln(\chi^2_{\rm LRLF}),
\end{equation}
where $S$ is the likelihood
function for  two-population models, analogous to $S$ defined in Eqn.~(\ref{eqn:maxlike}), $\chi^2_{\rm sc}$
is the $\chi^2$ value for the radio source counts and $\chi^2_{\rm LRLF}$ 
is the $\chi^2$ value for the local RLF.
$N_{\rm A}, N_{\rm B}$ and $N_{\rm C}$ are the number of data
points associated with the 3CRR, 6CE and 7CRS $L_{\rm [OIII]}$ and $L_{151}$ data ($n=302$ objects), 
the binned source count data ($30$ 6C and $16$ 7C bins) 
and the local RLF constraint ($1$) respectively.
The $\chi^2$ is given by

\begin{equation}
\chi^2 = \sum_{i=1}^{i=\rm nbins} \left( \frac{f_{{\rm data},i} - f_{{\rm model}, i}}{\sigma_{{\rm data}, i}}  \right)^2,
\end{equation}

\noindent and for the source counts
 \begin{equation}
f_{{\rm data},i} = \left(\frac{{\rm d}N}{{\rm d}N_0}\right)_i , 
\end{equation}
where ${\rm d}N_0 = 2400(S_{\rm min}^{-1.5} - S_{\rm max}^{-1.5})$
is the Euclidean source counts.

\subsection{Results from the two-population GLFs}
\label{sec:2popres}

\begin{table*}
\begin{center}
\begin{tabular}{ccccccccccccccc}
\hline\hline
Model&$\log(\rho_{10})$&$\log(\rho_{20})$&$p_2$&$\alpha_{\rm cut,1}$&$\alpha_{\rm cut,2}$&$z_1$&$z_{2{\rm a}}$&$z_{2{\rm b}}$&$\beta_o$&$\log_{10}(\sigma_{\beta}^2)$&$k$&$\Theta_{01}$&$\Theta_{02}$&$P_{\rm 2RS}$\\
\hline\hline
2RS &$-4.486$&$-3.776$&$31.47$&$-0.0501$&$0.0249$
  & $0.437$&$1.684$&$0.447$&$-0.0219$&$-3.283$&$3.436$&$20.5$&$29.9$&$1.00$\\
2SS &$-4.486$&$-3.776$&$31.47$&$-0.0500$&$0.0249$
  & $0.437$&$1.684$&$0.447$&$-0.0219$&$-3.283$&$3.437$&$20.6$&$52.7$&$1.59$\\
2RR &$-4.486$&$-3.776$&$31.47$&$-0.0501$&$0.0249$
  & $0.437$&$1.684$&$0.447$&$-0.0219$&$-3.283$&$3.436$&$29.4$&$31.0$&$2.08$\\
\hline
2RR &$0.152$&$0.020$&$1.33$&$0.023$&$0.0015$
  & $0.045$&$0.044$&$0.032$&$0.0007$&$0.019$&$0.556$&$21.7$&$^{+22.9}_{-15.8}$&\\
2SS &$0.143$&$0.020$&$1.34$&$0.022$&$0.0015$
  & $0.035$&$0.043$&$0.032$&$0.0007$&$0.019$&$0.516$&$48.2$&$17.3$&\\
2RR&$0.152$&$0.020$&$1.34$&$0.024$&$0.0015$
  & $0.047$&$0.044$&$0.033$&$0.0007$&$0.019$&$0.563$&$^{+48.5}_{-25.5}$&$^{+24.4}_{-16.5}$&\\
\hline\hline

\end{tabular}
{\caption[]{\label{tab:param2}{Best-fit parameters
for the two-population models.
The upper section gives the optimized parameters and the
lower section gives the errors on these best-fit parameters.
$\Theta_{01}$ and $\Theta_{02}$ give the constant transition angles $\Theta_{\rm trans_1}$ and $\Theta_{\rm trans_2}$
for a simple unified scheme in 
population $1$ or $2$ i.e. population $1$ of 2RS and 2SS and population $2$ of 2SS. For receding-torus schemes,
$\Theta_{01}$ and $\Theta_{02}$ are related to $\Theta_{\rm trans_1}$ and $\Theta_{\rm trans_2}$ by Eqn.~\ref{eqn:thetatrans},
using the median $\log_{10}(L_0) = 35.405$. The values of $S_{\rm min}$,
where $S_{\rm min}$ is the minimum value of the likelihood function $S$, are 
$2933.78, 2933.91$ and $2934.10$ for 2RS, 2SS and 2RR respectively and $\det(\nabla \nabla S_{\rm 2RS} |_{S_{\rm min}}) = 2.12 \times 10^{43}$, 
$\det(\nabla \nabla S_{\rm 2SS} |_{S_{\rm min}}) = 7.42 \times 10^{42}$ and $\det(\nabla \nabla S_{\rm 2RR} |_{S_{\rm min}}) = 3.58 \times 10^{42}$.
Inserting these values into Equation \ref{eqn:rel}, we find
the probability relative to model 2RS, $P_{\rm 2RS}$.}}}
\end{center}
\normalsize
\end{table*}

The maximum-likelihood
routine, where $S$ is defined in a similar way to that in
Eqn.~\ref{eqn:maxlike}, was performed on the two-population
models, to find best-fitting values of
the GLF parameters, and the results are given in Table \ref{tab:param2}.
GLF 2RR, with a receding-torus model in both high- and low-$\alpha$
populations was the best-fitting model but, in terms of probability, it is only preferred over
2RS by a factor of $\simeq 2$
(and over 2SS by a factor $1.3$). 
The scatter about the $\log_{10}L_{\rm [OIII]} - \log_{10}L_{151}$ relation was found to be $0.62$.

\begin{figure}
\begin{center}
\setlength{\unitlength}{1mm}
\begin{picture}(150,60)
\put(0,-10){\includegraphics{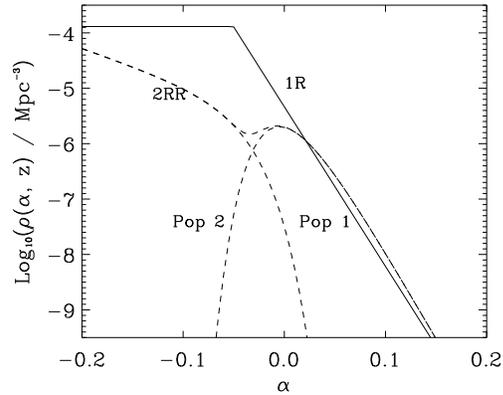}}
\end{picture}
\end{center}
{\caption[junk]{\label{fig:rhoalp}
{The GLF, $\rho(\alpha, \beta, z)$, integrated over
$\beta$ to produce
$\rho(\alpha, z)$ for $z = 1$.
The solid line is the GLF 1R and the dashed lines show the 2RR GLF.
The total GLF and the contributions from population $1$ and population $2$ are shown.}}}
\end{figure}

\begin{figure}
\begin{center}
\setlength{\unitlength}{1mm}
\begin{picture}(150,60)
\put(0,-10){\includegraphics{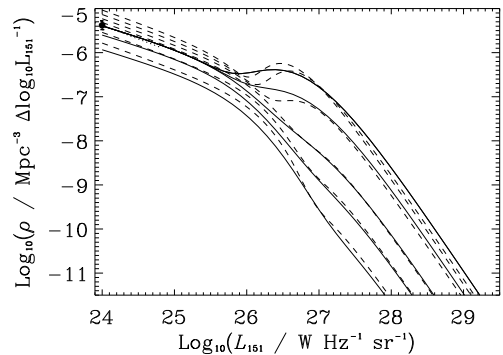}}
\end{picture}
\end{center}
{\caption[junk]{\label{fig:rlf2pop}
{The radio luminosity function at $151$ MHz, $\rho(L_{151}, z)$,  from
\citet{wrlf} (dashed lines) compared with the optimized
two-population GLFs for model 2RS (solid lines), (2SS and 2RR give
almost identical RLFs and are not plotted). The GLFs are plotted
at $z = 0.0001, 0.25, 0.5, 1.0, 2.0, 3.0$, and can be identified since
the lines corresponding to each redshift increase monotonically in
normalization with redshift. The dotted line indicates the fixed slope
adopted for the low-$\alpha$ population, and the filled circle indicates
the normalization of the local RLF at $z \simeq 0.25$ and $\log_{10}(L_{151} / \rm{W Hz^{-1} sr^{-1})} = 24.0$ from \citet{sad}.
Note the much less prominent `hump' artefacts in or new GLF-derived RLFs.}}}
\end{figure}

The form of the best-fitting GLF can be seen in  Fig.~\ref{fig:rhoalp}, which shows 
$\rho(\alpha, z)$ (the GLF integrated over $\beta$) against
$\alpha$. The two-population GLF is similar to the one-population GLF at
high $\alpha$, but they differ at low values of $\alpha$, where the 
two-population model is additionally constrained by radio source counts.
The RLF derived from the two-population GLF now agrees quite well with the Model
B RLF from \citet{wrlf} over the full range of $L_{151}$ and $z$ (Fig.~\ref{fig:rlf2pop}). 
\citet{sb} note that the two-population RLFs of \citet{wrlf} have prominent
humps due to the different evolutionary forms used for the high- and 
low-luminosity populations, which are identical in form to those used
for the high-and low-$\alpha$ populations in this study. 
They argue that FRI and FRII radio galaxies should not be treated as
intrinsically different classes of object, and that evolution should 
simply be a function of radio power, so that the higher power FRII objects
should undergo stronger evolution than FRIs. 
Since FRI objects are all drawn from the low-$\alpha$ population of the
GLFs defined in this study, their
evolution is in general weaker than FRIIs, which may be drawn from 
either population.
The RLFs derived from the GLFs 
are smoother than the RLFs found by \cite{wrlf} 
since at each value of $L_{151}$ objects are drawn from a band of objects
in $\alpha$. It seems that the RLFs derived from GLFs naturally produce smoother results
because the effects of scatter, if only partially, are taken into account.
The 2RS, 2SS and 2RR two-population models all give very similar GLFs,
so that the small likelihood differences probably arise from the differences in
their unified schemes.

The best-fitting unified scheme parameters are compared with the quasar
fraction as a function of $L_{\rm [OIII]}$ in Fig.~\ref{fig:fq} and
as a function of $\alpha$ in Fig.~\ref{fig:qfalpha}. The effect of
combining a simple unified scheme with a two-population model can
be seen, as the two-population GLF 2SS fits the data much better than the one-population
GLF 1S. The error bars on the quasar-fraction data are sufficiently
large that, despite the very different shape of the quasar fraction curves,
the non-luminosity-dependent schemes used in 2SS do not give a significantly worse fit
than GLFs employing a receding-torus model in one or both populations. 
These GLFs, 2RS and 2RR, fit the quasar fraction data very well, differing only at low values of
$\alpha$, where 2RR predicts that the quasar fraction goes to zero at very low $\alpha$. 
In comparison, 2RS and 2SS seem to over-predict slightly the number of objects at low $\alpha$,
in order to be able to account for objects like 3C386. It is obvious in this context why there
is little difference between the likelihoods of the GLFs. 2RR can fit the second lowest
data point in  Fig.~\ref{fig:fq} and Fig.~\ref{fig:qfalpha}, whereas 2RS and 2SS can both fit
the lowest data point. 
None of the GLFs can fit the third lowest quasar fraction data point in either Fig.~\ref{fig:fq}
or Fig.~\ref{fig:qfalpha}, which seems to arise from the presence of weak quasars at approximately
$-0.1 < \alpha < -0.3$ (see Fig.~\ref{fig:pca}).

\begin{figure}
\begin{center}
\setlength{\unitlength}{1mm}
\begin{picture}(150,60)
\put(0,-10){\includegraphics{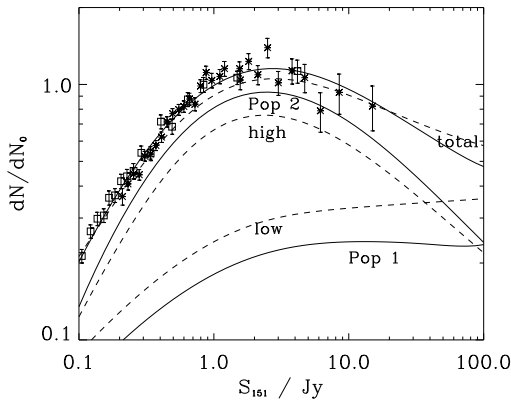}}
\end{picture}
\end{center}
{\caption[junk]{\label{fig:sc}
{The source counts from the 6C catalogue (stars) and 7C (squares), with the integrated source counts from the \citet{wrlf} RLF with contributions from the high and low populations (dashed lines), and from model 2RS (solid lines) with the contributions from population $1$ and population $2$ AGN shown separately.
}}}
\end{figure}

\begin{figure}
\begin{center}
\setlength{\unitlength}{1mm}
\begin{picture}(150,60)
\put(0,-10){\includegraphics{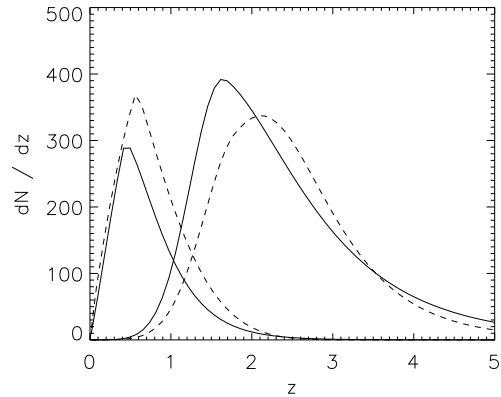}}
\end{picture}
\end{center}
{\caption[junk]{\label{fig:nofz}
{The redshift distribution $\rm{d}N/\rm{d}z$ at a flux-density limit of $0.1$ Jy for both populations of model 2RS (solid lines) and of Model B RLF from \citet{wrlf} (dashed lines). The plots are normalized such that the total number of 
sources in the sample is $1000$.}}}
\end{figure}

The source counts derived from the GLFs are compared with the 6C and 7C source count data 
(Sec.~\ref{sec:scdata}) and the
source counts derived from the RLF from \citet{wrlf} in Fig.~\ref{fig:sc}.
The RLFs from \citet{wrlf} do not give a very good fit to the source count data (reduced $\chi^2 = 3.17$),
under-producing the 6C counts at $\simeq 0.7-3.0$ Jy.
However, the employment of the Bayesian `hyper-parameters'
in the definition of the likelihood function (Eqn.~\ref{eqn:hyper})
leads to a sufficiently strong source count constraint, so that the source counts
derived from our GLFs result in an improved fit to the data (reduced $\chi^2 = 1.68$).
The difference between the data and the RLFs can be accounted for by
shot noise and by deviations of the number of objects in the 7CRS sample
from the average number of sources in an area of sky due to large-scale
structure, (e.g. \citealt{brand}).
Another consequence of the tighter source count constraint is that,
despite the good agreement of the RLFs at most redshifts 
(Fig.~\ref{fig:rlf2pop}), the low-$\alpha$
population $1$ provides less of the source counts than the low-luminosity
population of \citet{wrlf}, and correspondingly the high-$\alpha$ 
population $2$
provides more of the source counts than the \citet{wrlf} high-luminosity 
population, (which is also evident in Fig~\ref{fig:nofz}).

These subtle changes can have significant impact on the redshift
distributions predicted for new
radio source redshift surveys. For example, 
the peak of the model redshift distribution is shifted to lower 
redshifts for the GLFs compared to the \cite{wrlf} RLFs at the flux density
limit of a redshift survey at $0.1 ~ \rm Jy$ (Fig~\ref{fig:nofz}).

\begin{figure*}
\begin{center}
\vspace{6cm}
\begin{tabular}{cc}
\vspace{5cm}
\begin{picture}(230,60)
\put(-5,-10){\includegraphics{data.ps}}
\end{picture}
&
\begin{picture}(230,60)
\put(-5,-10){\includegraphics{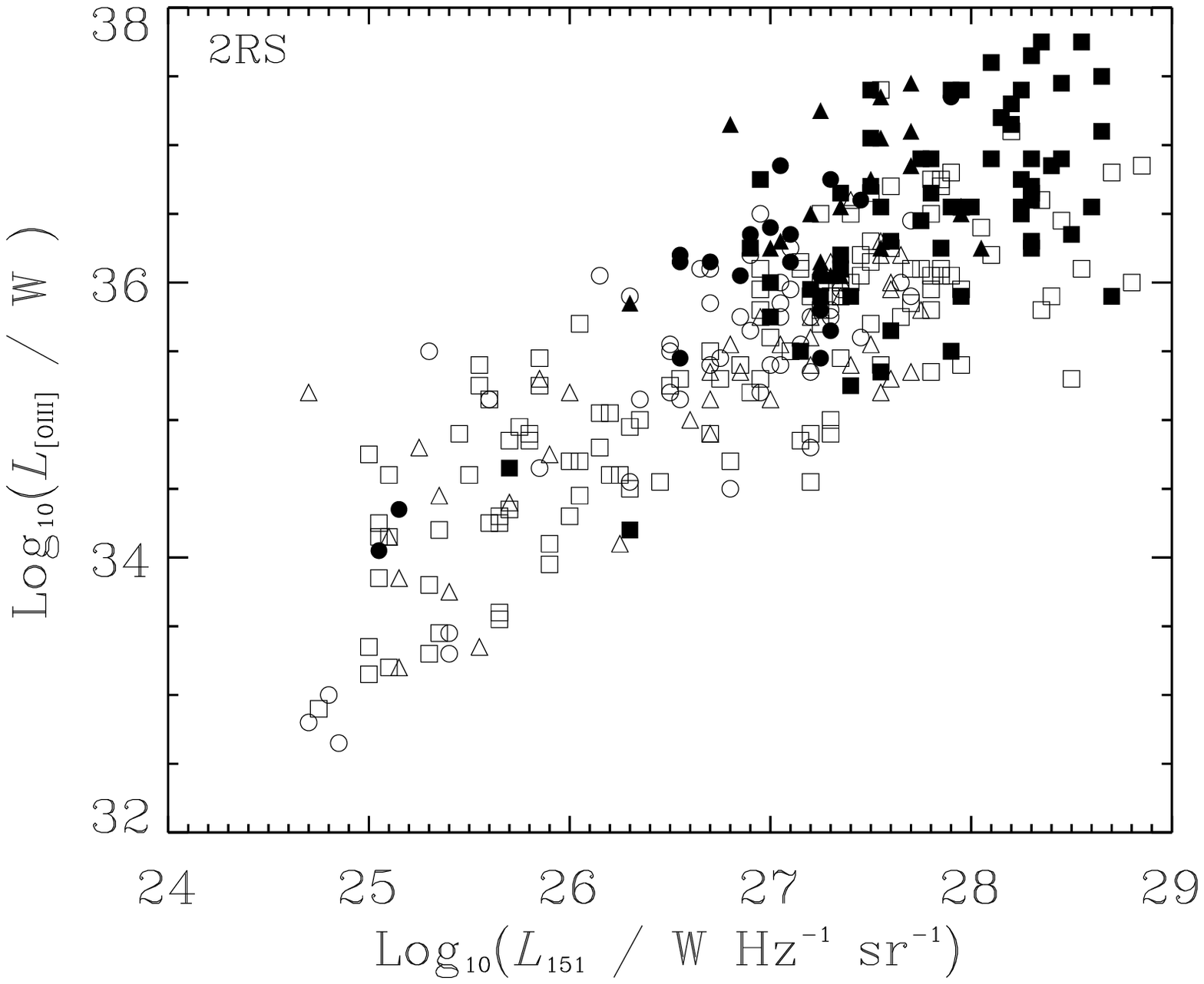}}
\end{picture}
\\
\begin{picture}(230,60)
\put(-5,-10){\includegraphics{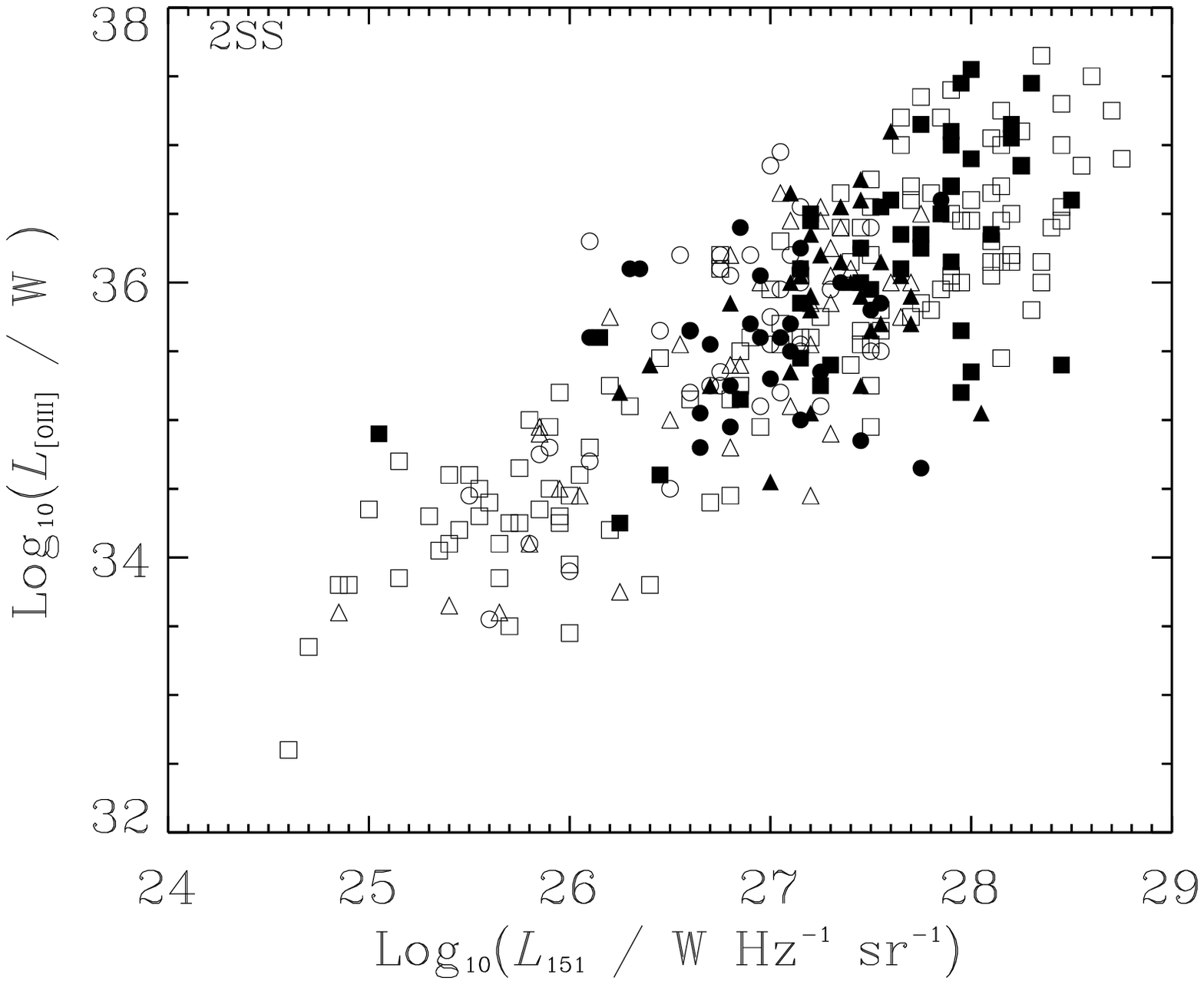}}
\end{picture}
&
\begin{picture}(230,60)
\put(-5,-10){\includegraphics{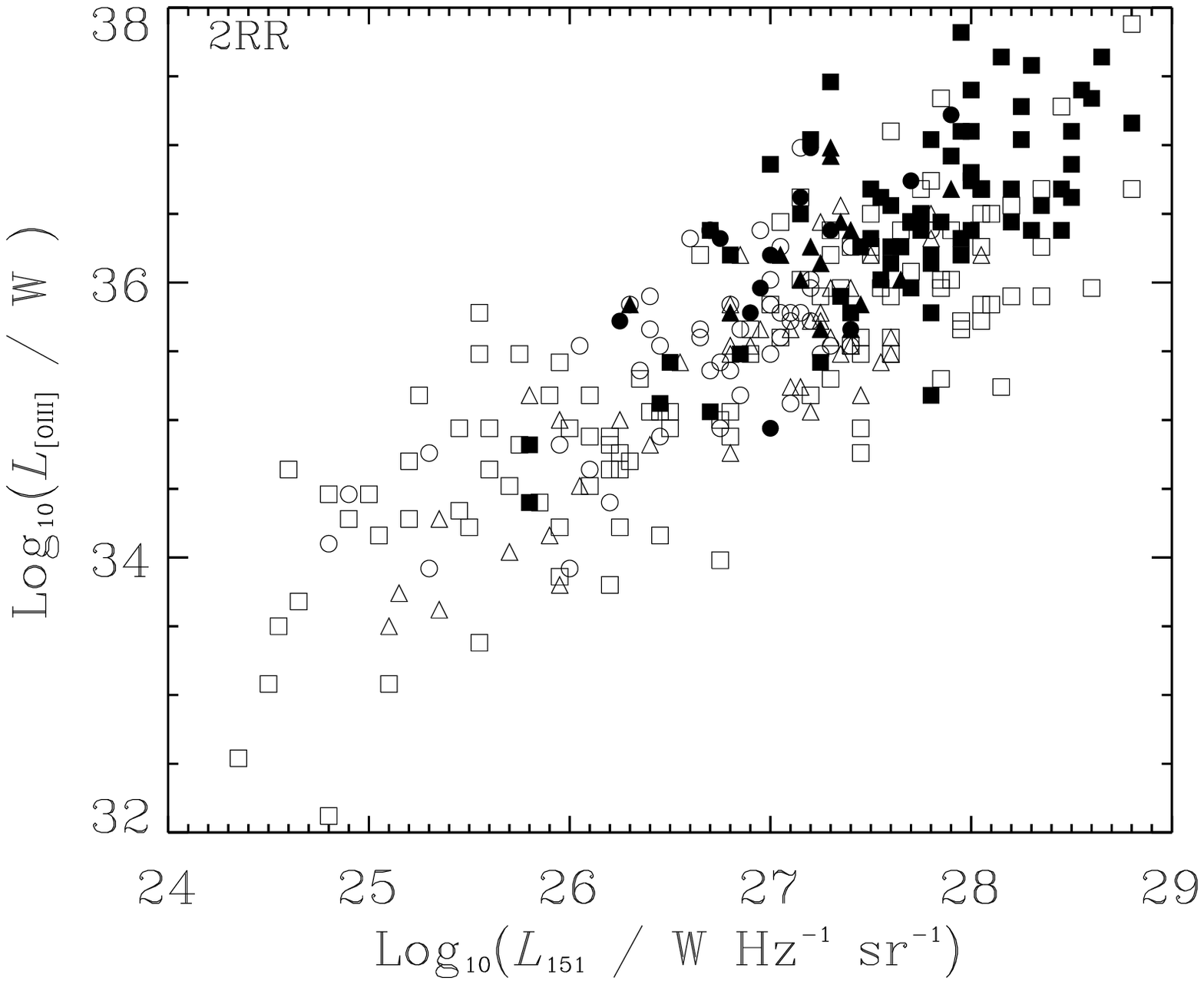}}
\end{picture}
\\
\end{tabular}
\end{center}
{\caption[junk]{\label{fig:sim_cut}
{The [OIII] emission-line luminosity against the $151$ MHz luminosity data and
simulations of this data
from the two-population GLFs for models 2RS, 2SS and 2RS.
Symbols are as in Fig.~\ref{fig:pz}.}
}}
\end{figure*}

The simulations of the data for the two-population
GLFs are shown in Fig.~\ref{fig:sim_cut}. All three GLFs show that
radio quasars are more likely to be found at high values of $L_{\rm [OIII]}$
compared to radio galaxies, although the distribution of quasars in the 
high-$\alpha$ population is more evenly distributed in 2SS because of the
constant opening angle of the torus. 
The most obvious problem with the simulations is that all of the GLFs under-produce the number
of objects at the lowest radio and emission line luminosities. This is a consequence of the
reduced normalization of population $1$ compared to the \citet{wrlf} RLF at 3CRR flux densities
(see Fig.~\ref{fig:sc} and Fig~\ref{fig:nofz}). 
This could be fixed by invoking a more complicated model.

\section{Discussion}
\label{sec:discussion}

\subsection{Quantitative Comparison of Unified Schemes}
A preliminary investigation (Sec.~\ref{sec:model}) showed that a 
receding-torus scheme is
strongly preferred over a simple non-luminosity-dependent unified
scheme. However when a two-population model was investigated, there
was little evidence to favour a receding torus 
in the high-$\alpha$ population. It is clear that, with enough scatter
in the radio - optical relation, a small quasar fraction in the 
low-$\alpha$ population 
combined with even a constant value of $\theta_{\rm trans}$ in the high-$\alpha$
population will mimic the rise of quasar fraction with emission-line luminosity
predicted by a luminosity-dependent
unified scheme such as the receding-torus model. 
A simple unified scheme alone will not generate the observed increase
in quasar fraction with $\alpha$ and the emission-line differences
between radio quasars and radio galaxies
but a receding-torus model or a two-population scheme with any
sort of unified scheme in the high-$\alpha$ population will produce
these effects. 
On the basis of quasar fractions and emission-line differences alone, 
there is not enough evidence to distinguish 
between possible luminosity dependences of the unified schemes, apart
from ruling out schemes with no luminosity dependence, such as model 1S.

\citet{sconf} reviews the evidence for the receding-torus model
highlighting the quasar fraction and emission-line difference
arguments.
However we have shown here quantitatively that the quasar fraction and the
differences in emission-line luminosity in radio-selected samples
are not very significant either individually or when considered together.

\begin{figure}
\begin{center}
\setlength{\unitlength}{1mm}
\begin{picture}(150,60)
\put(0,-10){\includegraphics{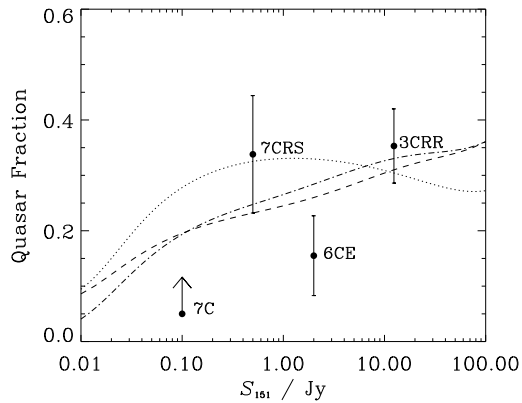}}
\end{picture}
\end{center}
{\caption[junk]{\label{fig:qfs151}
{The quasar fraction against the flux density at $151$ MHz $S_{151}$ from model 2SS (dotted), 2RS (dashed) and 2RR (dot-dashed). Also shown are the quasar fractions at the flux-density limits from the 3CRR, 6CE and 7CRS radio-selected 
complete samples used as constraints in the modelling. The lower limit on the quasar fraction 
from the 7CQ sample \citep{riley} is also shown. The error bars are the $\sqrt N$ errors.}}}
\end{figure}

A major cause of the inability of the data to discriminate between the competing
models is that the quasar fractions of the 
6CE sample is very different ($15.5 \%$) to the 3CRR ($34.7 \%$) and 7CRS samples
($33.8 \%$).
Fig.~\ref{fig:qfs151} shows the quasar fractions as a function of the
limiting flux density for the 3CRR, 6CE, 7CRS and 7CQ \citep{riley} samples and the
curves predicted from the 2RS, 2SS and 2RR models. This illustrates
the fact that although the 2SS model leads to very different
predictions than the receding-torus model, they cannot yet be clearly distinguished
by the data.

\begin{figure}
\begin{center}
\setlength{\unitlength}{1mm}
\begin{picture}(100,120)
\put(0,-20){\includegraphics{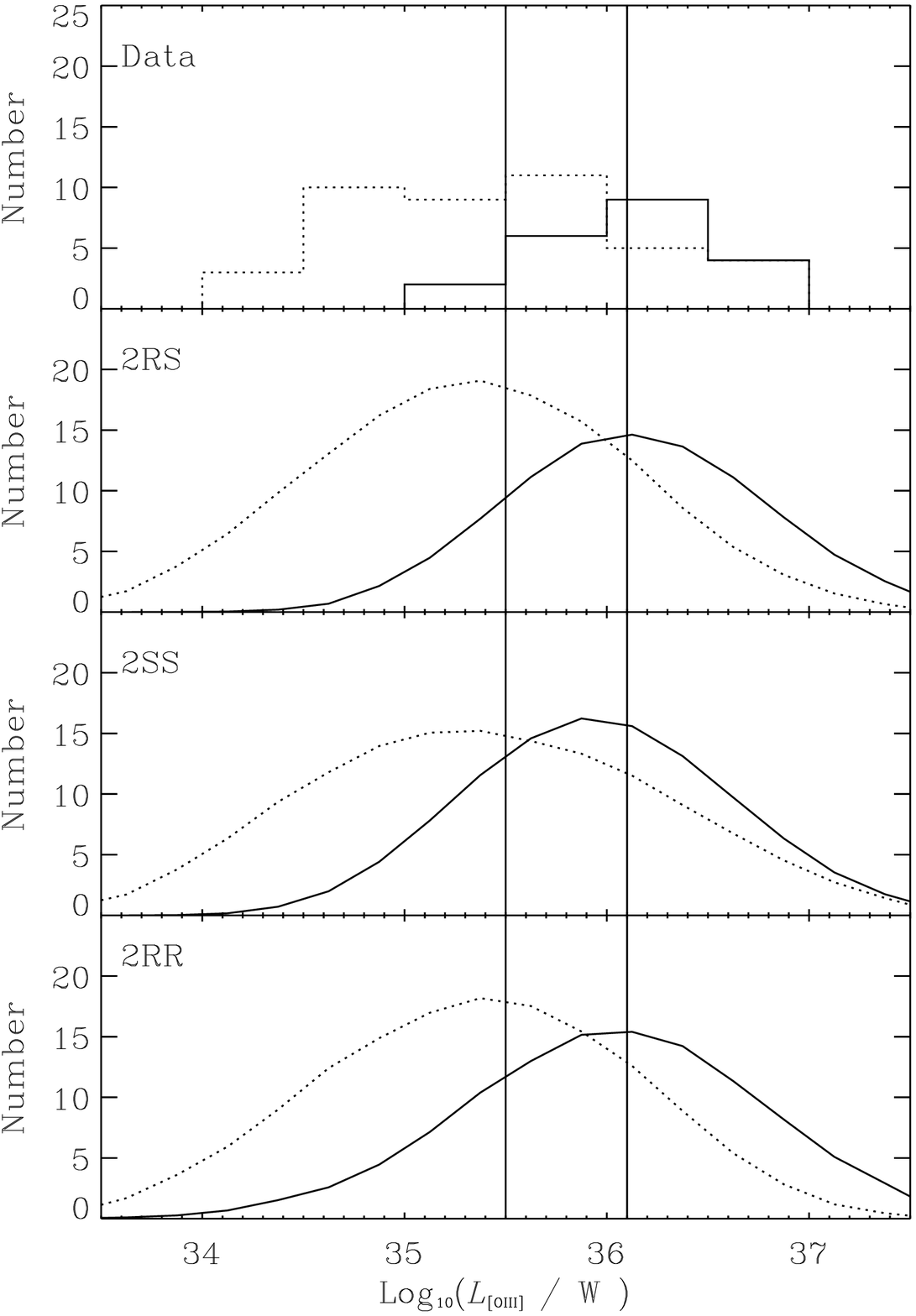}}
\end{picture}
\end{center}
{\caption[junk]{\label{fig:o3dist}
{
The top panel shows the number of 3CRR radio quasars (solid lines)
and radio galaxies (dotted lines) at $0.2 \le z \le 0.8$ at a
given $L_{\rm [OIII]}$ luminosity. The other panels show the
distribution of 3CRR radio quasars (solid lines) and radio galaxies
(dotted lines) at the same redshifts drawn from simulations
of the 2RS, 2SS and 2RR GLFs. The vertical solid lines show the positions of the median
value of $L_{\rm [OIII]}$ for radio galaxies and radio quasars.
}}}
\end{figure}

To re-investigate the issue of the emission-line differences between radio quasars
and radio galaxies, we repeat the experiment
of \citet{jb}. They compared each of $12$ 3C radio galaxies with 
radio quasars with $0.19 < z <0.85$, matched in redshift to within $0.05$ and 
in luminosity to within $30$ per cent, and find that quasar ${\rm [OIII]}$ 
luminosities exceed those of radio galaxies by a factor of $5-10$.
We plot histograms of the number
of 3CRR radio quasars and radio galaxies with $0.2 \le z \le 0.8$ 
in $L_{\rm [OIII]}$ bins, Fig.~\ref{fig:o3dist}. There is clearly
an offset in the distributions, with the median value of $L_{\rm [OIII]}$
being $36.10$ for quasars and $35.50$ for radio galaxies, giving
a shift of $0.60$, so that the median quasar has $4$ times more
luminous [OIII] emission than the median radio galaxy.
We found the distributions in $L_{\rm [OIII]}$ of 3CRR objects in this redshift 
range for the three two-population models. All three models
reproduce a shift in the median value of $L_{\rm [OIII]}$ between
quasars and radio galaxies. The 2RS model gave a shift of $0.78$ 
compared to $0.56$ for 2SS and $0.64$ for 2RR. 
This analysis shows that
any of the two-population models considered here can reproduce
the results of \citet{jb} so that their test does
not provide clear support for the receding-torus model. 
All models, with or without a receding torus, predict that
quasars are brighter than radio galaxies because the probability
of an object being classified as a quasar increases with $\alpha$ and $L_{\rm [OIII]}$.

\subsection{The meaning of $\alpha$ and $\beta$}

It is instructive to consider, if only briefly, possible interpretations for the axes found
by the principal components analysis in the context of the receding-torus
scheme and the two-population scheme. Since $\alpha$ accounts for the
overwhelming majority of the scatter, it must be a manifestation of
some important physical process. Higher values of $\alpha$
mean higher low-frequency radio luminosity and higher emission-line luminosity,
i.e. more powerful objects. Emission-line luminosity is often assumed to
be dominated by photo-ionization from a strong nucleus, though there
is evidence that some part, specifically low-ionization lines like 
$\rm{[OII]}$, is powered by radiative shocks \citep{inskip}. 
In this context an increase
in accretion rate will cause an increase in both radio luminosity
and emission-line luminosity \citep{rs91}, although it is not immediately obvious in
what proportions. We tentatively suggest that
$\alpha$ is an indicator of accretion rate and $\beta$ is the scatter about this
relation, which can probably be attributed to a number of factors:
for example, environment and black hole mass. In this context and for
the two-population
model, it seems possible that $\alpha_{\rm cut,1}$ and $\alpha_{\rm cut,2}$
represent respectively a critical accretion rate above which AGN are shining at
a significant fraction of their Eddington luminosity, and a critical rate 
below which they are shining at a tiny fraction. 

In essence, it can be shown that, for the very similar values of scatter in
the radio and emission-line distributions as found for this sample,

\begin{equation}
\alpha \propto \log_{10} \left( \frac{ L_{151} L_{\rm [OIII]} }{ L_{151,{\rm av}} L_{\rm [OIII], av} } \right),
\end{equation}

\noindent and

\begin{equation}
\beta \propto \log_{10} \left( \frac{ L_{151}  / L_{\rm [OIII]} }{ L_{151, {\rm av}} / L_{\rm [OIII], av} } \right).
\end{equation}

$L_{151}$ gives an indication of the jet power which is in turn powered
by the central engine and $L_{\rm [OIII]}$, arising from photo-ionization,
is also a measure of the power of the central engine. 
$\alpha$ represents the product of the radio (mechanical) and optical
(radiative) output, compared
to the average product. $\beta$ represents the ratio of radio output
to optical output compared to the average ratio.

As $\beta$ probably represents the scatter from a number of different sources,
it would be interesting to see if $\beta$ correlates with any of the most likely 
contributions towards scatter. 
It has been shown by \citet{mag} that the black hole mass
correlates with the mass of the bulge in galaxies, and assuming a constant
mass-to-light ratio, we expect the stellar luminosity to correlate with
the black hole mass.
\citet{wkz} find that the best-fit $K-z$ relation from the
3CRR, 6CE, 7CRS and 6C$^{\ast}$ samples is given by

\begin{equation}
\label{eqn:K}
K = 17.37 + 4.53 \log_{10} z - 0.31 (\log_{10} z)^2.
\end{equation}

\begin{figure}
\begin{center}
\setlength{\unitlength}{1mm}
\begin{picture}(150,60)
\put(0,-10){\includegraphics{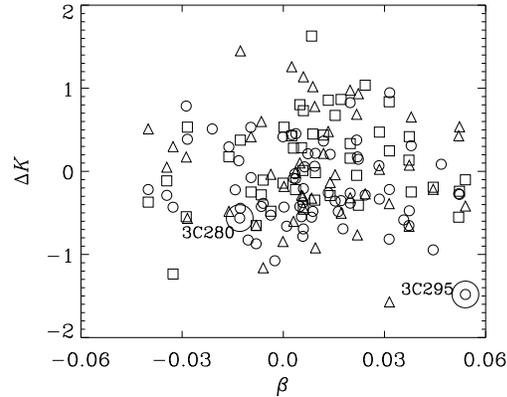}}
\end{picture}
\end{center}
{\caption[junk]{\label{fig:betaK}
{The excess $K$ magnitudes over the best fitting $K-z$ relation of Eqn.~\ref{eqn:K}, $\Delta K$, from
\citet{wkz},
against $\beta$ for radio galaxies from the 3CRR, 6CE and 7CRS (I and II)
samples, with symbols as in Fig.~\ref{fig:pz}. Large circles surrounding two 3C objects
indicate objects known to be in clusters \citep{hardcastle}.}}}
\end{figure}

In Fig.~\ref{fig:betaK} a measure of the stellar luminosity $\Delta K$
(the excess $K$ luminosity over the $K-z$ relation of Eqn.~\ref{eqn:K}
\footnote{\citet{wkz} show that the empirical $K-z$ relation of Eqn.~\ref{eqn:K}
lies very close to the locus expected for a giant elliptical of fixed mass whose
stellar population formed at high redshift and is evolving passively.
$\Delta K$ represents (in magnitudes) the excess luminosity.})
is plotted
against $\beta$ for $160$  radio galaxies from the 3CRR, 6CE and 7CRS 
(I and II) samples \citep{wkz}. 
The Spearman rank correlation coefficient is $-0.064$ with significance $0.42$
implying that $\beta$ is not significantly correlated with black hole mass. 
The correlation coefficient between $\alpha$ and K is  $-0.131$ with
significance $0.10$.

\citet{lacy} show that, for a sample of quasars from the FIRST Bright 
Quasar Survey (FBQS) and optically-selected quasars from the Palomar-Green survey, $L_{5 \rm GHz} \propto M_{\rm BH}^{1.9 \pm 0.2} (L/L_{{\rm Edd}})^{1.0}$, where $M_{\rm BH}$ is the black hole mass.
This agrees with the slope of $1.05$ found here in the $\log_{10}L_{151} -
\log_{10}L_{\rm [OIII]}$ relation if the narrow emission-line 
luminosity, $L_{{\rm [OIII]}}$ is proportional to the bolometric 
luminosity, $L$, and if the range in black-hole mass is small.
We conclude that the scatter $\sigma_{\beta}$ is not dominated by the
black-hole mass.

There are of course a variety of other factors which might correlate with 
$\beta$. For example, implicit in the work of \citet{rs91} is that $\beta$
might be expected to correlate positively with the richness of the 
radio source environment as, for a given jet power (assumed to scale
with bolometric quasar luminosity), simple
physical models predict that radio sources have higher radio luminosities
in dense-gas environments (see also \citealt{ba96}), leading to more positive
values of $\beta$.
Sadly, despite the existence of XMM-Newton and Chandra, there are as yet no
comprehensive studies of radio source environments that can be used to make 
such a test. We label on Fig.~\ref{fig:betaK}, several radio sources known to 
be in rich galaxy clusters from ROSAT observations \citep{hardcastle}. 
The radio galaxy 3C295 ($\beta = 0.054$) is one of the highest-$\beta$ radio 
galaxies in the sample, in agreement with the prediction, but
3C220.1 (which is not in Fig.~\ref{fig:betaK} as it does not have a measured 
K magnitude) has $\beta = 0.015$ and 3C280 has $\beta = -0.013$,  compared to 
the average value of $\beta = 0.0025$ for radio galaxies in the redshift 
range $0.4 < z < 1$. \citet{hardcastle}
also finds that five 3CRR quasars with $0.35 < z < 0.75$ reside in rich cluster
environments, and they have an average $\beta = -0.0024$, compared to an 
average of $-0.0037$ for quasars in this redshift range. So there are
hints that this correlation may be present, but obviously the numbers involved 
are very small. 
We have also failed to find any significant correlation 
between $\beta$ and projected radio source 
linear size $D$. Naively one might expect
a negative correlation because synchrotron and adiabatic losses
should cause a systematic decline in radio luminosity over the lifetime
of the radio source if the jet power and narrow-line
luminosity remain fixed \citep{kda,brw}. However, as discussed by \citet{wem}, 
such a correlation may be masked by a number of effects including
changes with time of jet power and $L_{\rm phot}$, and the boosting of
emission-line fluxes in small sources due to radiative 
bow shocks \citep{inskip}.

\section{Conclusions}
A new approach to investigating unified schemes has been presented, 
based on a principal components analysis of the 3CRR, 6CE and 7CRS 
complete samples. Generalized luminosity functions have been derived
based on these principal components: $\alpha$ which encodes the 
$L_{151} - L_{\rm [OIII]}$ correlation and $\beta$ which encodes the
scatter about this correlation. The main advantage of this new approach
has been that the unified scheme parameters have been found by taking
into account the intrinsic scatter in the $L_{151} - L_{\rm [OIII]}$
correlation. The main conclusions to be drawn from
this analysis are as follows.

\begin{enumerate}
\item A receding-torus model is strongly favoured  over a simple
non-luminosity-dependent unified scheme for GLFs with one population
of radio sources.

\item With the extra constraint of 6C and 7C radio source counts and
the normalization of the local RLF, two-population GLFs were derived.
The GLFs give rise to RLFs which
are very similar but smoother than those of \citet{wrlf}. 
There is very little difference in likelihoods between a GLF
with a receding torus in both populations, a GLF with a torus opening angle that does not
vary with ionizing luminosity in both populations, and a GLF with a receding-torus
in the high-$\alpha$ population and a constant-opening-angle torus in the low-$\alpha$
population.

\item Two-population models reproduce the radio survey data well, and
can provide a natural explanation for the rise in quasar fraction with
emission-line luminosity and the emission-line differences between 
radio quasars and radio galaxies. 
The receding-torus may be a feature in both populations but
this is not yet proved. 
\end{enumerate}

\section*{Acknowledgements}
JAG acknowledges the support of a graduate studentship from Oxford 
University and Magdalen College. SR acknowledges the 
support of a PPARC Senior Fellowship.
This research has made use of the NASA/IPAC Extragalactic Database, which
is operated by the Jet Propulsion Laboratory, Caltech, under contract
with the National Aeronautics and Space Administration.
We would like to thank Chris Simpson for a critical
reading of an early version of the manuscript, and an
anonymous referee for some very useful comments.

\label{lastpage}

\end{document}